**Article title: Development of computational models for emotional diary text analysis to support maternal care**

**Corresponding author: Lauri Lahti (email: lauri.lahti@aalto.fi)**

This article manuscript version was completed on 17 October 2017, and it was self-archived on the arXiv repository (https://arxiv.org/) on 17 October 2017.

**Manuscript length:** About 12 000 words. **Number of tables and figures:** 7+2.

# Development of computational models for emotional diary text analysis to support maternal care

**Lauri Lahti[1*], Henni Tenhunen[2], Seppo Heinonen[3], Minna Helkavaara[4], Maritta Pöyhönen-Alho[5], Paulus Torkki[2]**

1. Aalto University School of Science, Department of Computer Science, Helsinki, Finland

2. Aalto University School of Science, Institute of Healthcare Engineering, Management and Architecture (HEMA), Helsinki, Finland

3. Department of Obstetrics and Gynecology, Helsinki University Hospital and University of Helsinki, Helsinki, Finland

4. Department of Psychiatry, Helsinki University Hospital and University of Helsinki, Helsinki, Finland

5. Department of Obstetrics and Gynecology, Helsinki University Hospital and University of Helsinki, Helsinki, Finland

**Correspondence:**

Lauri Lahti

lauri.lahti@aalto.fi

**Abstract**

We propose new computational models for analyzing self-reported emotional diary texts of pregnant women to support maternal care. We gathered affective ratings outside clinical setting and developed new models to facilitate interpretation and communication of affective expressions between persons representing different affective ratings. Relying on constructed emotion theory, models of dimensional emotion categories and affective ratings of Self Assessment Manikin, we demonstrate our new proposal to analyze linguistic data with computational models exploiting vector space and clustering methods. 35 persons having Finnish as a native language provided affective ratings for 195 emotional adjectives and 16 pregnancy-related nouns in Finnish in dimensions of pleasure, arousal and dominance. We developed new models to represent dependencies and differences of affective ratings between various population subgroup categorizations, including "women without children", "women with children" and "men without children" that we consider important population segments to be addressed in maternal care.

Our affective ratings showed significant correlations between pleasure and dominance (like Warriner et al., 2013) and with previous data collections (Söderholm et al., 2013; Eilola & Havelka, 2010; Warriner et al., 2013). Our affective ratings had significant effects on categorizations based on gender, gender-parental role and the time of the day and duration of giving ratings. Our results indicate accordance with significant affectivity differences of gender and age (Warriner et al., 2013) and motherhood (Rosebrock et al., 2015). Our proposed models aim to support health-related communication. Our results suggest gathering next the affective ratings of patients of maternal care in a real clinical setting.

# 1. Introduction

## 1.1 Motivation for the current research

The patient's perspective towards the healthcare services and his/her choices of everyday life affecting the state of health depend largely on the motivation and preferences of the person which are based on emotions and affectivity[1]. Thus there is a strong need for understanding better and more reliably how the patient's emotions have an influence on the quality and value of healthcare services experienced and gained by the patient. By getting an increased understanding about the role and properties of emotions in the patient's life it is possible to develop methods for segmentation of patient groups and thus to identify and address their needs of tailored appropriate care.

To increase understanding about emotions in healthcare there is a need for gathering empirically quantitative data about affectivity in relation to diverse experiences by varied population groups. In healthcare context maternal care is a domain of life highly associated with concerns of the future and affectively sensitive events. These specific conditions of maternal care provide a valuable opportunity to carry out empirical research about emotional processes of the patient and their relation to health. Also in the context of maternal care there is a need for new methods helping to identify and address various personal, cultural and language-specific characteristics of experiences and linguistic expressions of affectivity of the patient.

Despite of increased knowledge about psychological properties of emotional processes, modeling the use of natural language and technological opportunities to gather and analyze health-related data, these resources have not yet been fully exploited to implement personalized healthcare. Therefore it is important to develop new mobile methods and tools that enable the patient to collect and report personal data about everyday life events for the healthcare system. Furthermore there is a need for developing methods to support reliable interpretation and communication of correct emotional information between persons who represent different representations of affectivity of emotional expressions and participate in the healthcare process of the patient. In respect to measuring experiences expressed in natural language it seems useful to create methods that rely on the affective ratings given for an appropriate set of emotional adjectives as well as words related to the current care.

A primary aim for our research was to develop tailored computational models for measuring,

---

1    In this article we use terminology of affective and emotional properties relatively interchangeably as done in previous research (Rennung & Göritz 2015).

interpreting and communicating affectivity in Finnish language usage to address specific population segments and contexts of maternal care. To accomplish this aim a supplementary aim for our research was to gather experimentally affective ratings concerning an appropriate set of Finnish emotional adjectives and pregnancy-related nouns from persons representing suitable complementing population segments. A further aim for our research was to analyze the gathered affective ratings with vector space and clustering methods and to represent dependencies and differences of affective ratings between various population subgroup categorizations. We wanted to exploit such categorizations that enable identification of properties of affectivity that have a risk to become misinterpreted and thus need support in communication when implementing a successful healthcare process.

On a practical level, an aim of our research is to evaluate how belonging to a specific subgroup of persons can have an effect on the type of emotions (affectivity) a person associates with certain concepts. Thus we aim to identify if persons belonging to different subgroups associate differently the affectivity measures for a set of concepts and what kind of patterns and dependencies can be found for these differences. Based on the mentioned aims we propose new computational models to support maternal care by analyzing *self-reported emotional diary texts of pregnant women*. We introduce a set of new models developed to support interpretation and communication of affective expressions between persons who represent different opinions concerning experiencing and characterizing affectivity of concepts.

By developing our models we aim to enable finding favorable ways to identify and support pregnant women that may have health problems such as suffering from mood disorders, psychosocial stress and antenatal anxiety. We suggest that our models can be exploited for developing computer-assisted methods to help both the pregnant woman and other persons participating in the maternal care process to understand better the properties of affective experiences and their dependencies to medical risks and conditions as well as to manage beneficially affectivity with emotion regulation strategies. Our these considerations are motivated by the previous research (Lee et al., 2016; Bos et al., 2013; Nie et al., 2017; Rosebrock et al., 2015; Edwards et al., 2017; Lever Taylor et al., 2016; Posner et al., 2005; Feldman, 1995).

## 1.2 Previous research

We next provide a review of previous research that gives motivation for our proposed methods of analyzing self-reported emotional diary texts of pregnant women.

Research concerning sociological and linguistic perspectives of healthcare has traditionally emphasized the interaction between a doctor and a patient and a lot of research has been carried out by people not directly involved in the healthcare work and with a limited emphasis on patient engagement (Candlin & Candlin, 2003; Elwyn, 2001). Analyzing the use of natural language to carry out social actions enables discovering a natural repeatable organization of structures and rules emerging in interaction and processes (Drew et al., 2001).

With the rising technological opportunities to record, share and access health-related data that can be increasingly collected by the patients themselves in everyday life with smart phones and mobile health-tracking devices it is important to emphasize research efforts for developing innovative evidence-based methods and tools to support data analysis in a real-time healthcare setting (Lahti, 2016). Previous results of mobile health experiments indicate that pregnant women may be an especially responsive population group for using web-based applications (Cormick et al., 2012; Parker, Dmitrieva, Frolov & Gazmararian, 2012). It has been suggested that temporal changes in pregnancy-related healthcare can be usefully measured with ecological momentary assessments (Bolger et al., 2003).

In respect to analyzing informal patient-driven self-reporting and health-related communication in a context dealing with emotionally loaded topics there are promising efforts of computational linguistics. These include research about transcribed communication of telephone health advice service of NHS Direct Corpus (Adolph et al., 2004) and online health advice services of Teenage Health Freak corpus and Lucy Anwers corpus (Harvey et al., 2013). In respect to analyzing unconstrained language of personal health-related communication there are promising research efforts. These include analysis of data retrieved from Twitter messages, including extraction of public health topics, public health trends and personal health experiences (Paul & Dredze, 2012; Parker, Wei, Yates, Frieder & Goharian, 2013; Jiang et al., 2016).

During pregnancy psychosocial stress is an important risk factor for preterm birth (Wadhwa et al., 2011) and antenatal anxiety can have adverse effects on the child's development (Van den Bergh & Marcoen, 2004). A systematic review of web-based interventions for the prevention and treatment of mood disorders in the perinatal period suggested that these interventions may improve maternal mood in post-partum period but more studies are needed especially with interventions delivered antenatally (Lee et al., 2016).

In the experiment of Bos et al. (2013) pregnant women were asked to rate their emotions with a Portuguese version of Profile of Mood States questionnaire (Azevedo et al., 1991; McNair et al., 1971) and it was found that negative affect was a predictor of postpartum depression whereas positive affect had a protective role, it was also suggested that the affectivity measurements showed agreement with the pleasure dimension and the activation (arousal) dimension of affect proposed by Russel (1980). The evaluation of Nie et al. (2017) concerning pregnant women threatened by a premature labor showed positive correlations between resilience, active coping, positive affect and social support whereas negative correlations were found between depression, pressure, passive coping, negative affect and social support as well as between depression, pressure and positive affect. In this evaluation of Nie et al. (2017) affectivity was measured with a Chinese version of Positive and Negative Affect Scale (PANAS) (Huang et al., 2003; Watson et al., 1988).

With unpleasant, neutral, pleasant and threat visual stimuli Rosebrock et al. (2015) showed that the pregnant and postpartum women and the age-matched women expressed lower arousal ratings relative to the college-aged women, and that postpartum women expressed increased arousal to unpleasant or threat stimuli compared to other stimuli. Challenges in maternal emotion regulation strategies can be indirectly related to higher infant negative affect (Edwards et al., 2017). For example mindfulness-based interventions in the perinatal period have been suggested to reduce maternal depression, anxiety and stress (Lever Taylor et al., 2016). It has been suggested that for some persons the overlapping cognitive schemas may decrease the ability to differentiate by arousal levels such emotions that have similar valence (pleasure) levels, for example emotions of sadness, anxiety, shame and fear (Posner et al., 2005; Feldman, 1995). This condition has been referred to as valence focus and has been supported by neuroimaging studies in respect to explaining the pathophysiology for depression and anxiety disorders as well as identified limited differentiation of affective states among children (Posner et al., 2005; Feldman, 1995).

According to the *word frequency effect* a person typically responds faster to high-frequency than low-frequency stimulus words of a language (Duyck et al., 2008) and according to the *age of acquisition effect* a person recognizes and produces faster stimulus words that he/she has learned earlier than later in the life (Izura & Ellis, 2002). It has been found that the emotional quality of a word can influence word recognition (Scott, 2014). In the case of high-frequency words the responses are faster for positive than negative or neutral words (Sereno et al., 2015). It has been suggested that faster access to positive information than negative information can be based on that positive information is more densely clustered in a semantic space (Unkelbach et al., 2008). It has been suggested that emotionally valenced words can have an advantage of facilitation in contrast with neutral words due to high arousal but since negative words have a low valence often representing a threat they are subsequently inhibited (Taylor, 1991; Pratto & John, 1991; Kuperman, 2015).

The neural networks of the brain can inherently fire without any external stimuli (Swanson, 2012) and it has been suggested that these *brain's simulations* can function as Bayesian filters for arriving sensory input and thus generate various psychological phenomena that include also emotions (Deneve,

2008; Bastos et al., 2012; Gallivan et al., 2016; Barrett, 2017). This assumption relies on a hypothesis referred to as predictive coding, active inference or belief propagation (Deneve & Jardri, 2016; Clark, 2013; Seth, 2013). According to a *constructed emotion theory* the brain builds meaning by predicting and adjusting to incoming sensations and the sensations are categorized so that they are both applicable situationally and thus meaningful in respect to previous experiences of the person (Karmiloff-Smith, 2009; Boiger & Mesquita, 2012; Xu & Kushnir, 2013; Bressler & Richter, 2015; Barrett et al., 2015; Barrett, 2017). Therefore if the previous experiences of a certain emotion are exploited to categorize the predicted sensory input and to conduct action then the person can be expected to experience this emotion in question (Barrett, 2017).

To identify and measure distinctiveness of some features in natural language the analysis can benefit from extending observation to supplementary dimensional and distributional projections and transformations of the original data set, including *vector space models*. A general assumption with vector space models is that the similarity or association of two items each represented with a vector can be evaluated based on the distance between these two vectors or based on the directional angle between them (for example based on a *cosine similarity measure*). Furthermore opposite vectors can be considered to represent items having opposite properties.

Previous research has suggested various emotional categorization approaches and two broadly accepted alternative viewpoints include *discrete emotion categories* (considering different emotions as relatively distinct types of experience) and *dimensional emotion categories* (considering different emotions to form a relatively smooth continuity of types of experience). A popular claim relying on discrete emotion categories is that formation of all human emotional experiences can be based on six basic emotions that are anger, disgust, fear, joy, sadness and surprise (Ekman et al., 1972). For dimensional emotion categories different scales have be proposed estimating for example the properties of the emotion in respect to pleasure and arousal or reaction to a given stimulus (Judge & Larsen, 2001). We decided to actively exploit dimensional emotion categories relying on Self Assessment Manikin scale (Bradley & Lang, 1999a; Lang et al., 2008) in our research since they offer practical opportunities for comparison of different affective ratings based on measures of distance and direction.

Previous affective ratings of native speakers of Finnish include collections of 440 Finnish nouns ($n_{persons\_giving\_ratings}$=996) (Söderholm et al., 2013) and 210 Finnish nouns ($n_{persons\_giving\_ratings}$=135) (Eilola & Havelka, 2010). It has been reported that for 54 shared words between the collections of Söderholm et al. (2013) and Eilola and Havelka (2010) there was a strong correlation in the valence ratings but not significant correlation in arousal ratings although in a pair-wise analysis carried out separately for negative, neutral and positive arousal ratings it was found that only negative nouns reached significant correlation in arousal ratings (Söderholm et al., 2013). By freely listing Finnish emotion words 2020 response items were collected ($n_{persons\_giving\_ratings}$=100) including 744 unique emotion words with some gender- and age-related differences (Tuovila, 2005).

### 1.3 Experimental evaluation in the Finnish context

We report an experimental evaluation of data acquisition and analysis methods carried out with 35 persons outside a clinical setting that enabled us to develop a set of new models to represent how the affective ratings can depend on various properties of language usage and population subgroup categorizations, including subgroups "women without children", "women with children" and "men without children" that we consider important population segments to be addressed in maternal care. Here the expression "with children" refers to having own child/children. We gathered affective rating data for emotional adjectives and pregnancy-related nouns in Finnish language and contrasted our data and results to other affective rating results and linguistic resources. With this experimental evaluation we wanted to examine the extent of needs for implementing new computational models for measuring, interpreting and communicating affectivity that can be tested and used in a real clinical setting.

Since our research is focused on the domain of Finnish public healthcare system and more specifically supporting maternal care we consider it important to address culturally dependent characteristics of communication and the use of Finnish language. To enable us to formulate a suitable experimental setup for evaluating affective ratings of an appropriate set of concepts of Finnish language concerning healthcare and maternal care settings we explored findings of the previous research. The properties of previous collections of emotion words and affective ratings in Finnish (Söderholm et al., 2013; Eilola & Havelka, 2010; Tuovila, 2005) provide a general foundation for our aim to develop a method of measuring the patient's experiences.

We aimed to gather new affectivity data of patients that is measured and categorized in a framework that has foundations in respected psychological theories, models and rating scales, such as constructed emotion theory (Barrett, 2017), models of dimensional emotion categories (Judge & Larsen, 2001) and affective ratings of Self Assessment Manikin (Bradley & Lang, 1999a; Lang et al., 2008). We aimed to implement our acquisition of affective rating data with a method addressing the previous notions of the responsiveness of pregnant women for using web-based applications (Cormick et al., 2012; Parker, Dmitrieva, Frolov & Gazmararian, 2012) and ecological momentary assessments (Bolger et al., 2003). For computational analysis of conceptual affectivity data and text-based patient report data there are various previous results that encourage the use of vector space and clustering methods to identify related patterns of data (Hassanpour & Langlotz, 2016; Toivonen et al., 2012).

## 2. Methods

### 2.1 Measures

Our experimental work relies on using *Self Assessment Manikin (SAM) test* that is based on dimensional emotion theories and assumes that differences in affectivity of words, objects and events can be represented with a three-dimensional measurement scale (Wundt, 1896; Osgood, 1952; Schlosberg, 1954; Mehrabian & Russell, 1974; Russell, 1980; Lang, 1980; Hodes, Cook & Lang, 1985).

Based on results of factor analysis it has been claimed that a three-dimensional measurement scale can possibly be a fundamental feature for the organization of human experiences semantically and affectively (Osgood, 1952; Mehrabian & Russell, 1974; Lang, 1980; Hodes, Cook & Lang, 1985). These three different dimensions of affectivity are often referred to in English as *pleasure* (or valence), *arousal* (or activation or alertness) and *dominance*. To enable an efficient way to define affectivity of a thing into three dimensions it has been developed a visual *SAM scale* (Lang, 1980; Hodes, Cook & Lang, 1985; Bradley & Lang, 1994) which generates three dimensions in factor analysis. In the visual SAM scale the dominance dimension is defined to be interpreted so that the lowest level of dominance (illustrated with the smallest SAM icon) means that the person feels being completely controlled whereas the highest level of dominance (illustrated with the biggest SAM icon) means that the person feels being completely in control (Bradley & Lang, 1999a).

The three-dimensional affectivity measurement model and SAM test are well respected frameworks with a long research tradition that has enabled creation of affective rating collections for various languages besides English (Bradley & Lang, 1999a) and Finnish (Söderholm et al., 2013; Eilola & Havelka, 2010), including German (Schmidtke et al., 2014), French (Monnier & Syssau, 2016), Italian (Montefinese et al., 2014), Spanish (Redondo et al., 2007). European Portuguese (Soares et al., 2012), Polish (Imbir, 2016) and Indonesian (Sianipar et al., 2016). Besides for words (Bradley & Lang, 1999a), affective ratings have been collected for texts (Bradley & Lang, 2007), pictures (Lang et al., 2008) and sounds (Bradley & Lang, 1999b).

It has been suggested that there is a relative independence of the three affectivity dimensions and on the other hand that there are some correlations between these dimensions (Mehrabian, 1972; Bradley & Lang, 1994; Bradley & Lang, 1999a). It has been suggested that the dimensions of valence and arousal have a quadratic relationship or a V-shaped relationship, valence has a positive linear relationship with dominance, and arousal and dominance have a quadratic relationship or a positive linear relationship (Bradley & Lang, 1999a; Warriner et al., 2013; Montefinese et al., 2014; Moors et al., 2013; Imbir, 2015; Kuppens et al., 2013). An extensive collection of affectivity measures in dimensions of pleasure, arousal and dominance for English language contains 13 915 English word lemmas, accumulated based on single dimension rating assignments with 1827 respondents (Warriner et al., 2013).

Using a SAM scale that relies on measuring a response to visual stimuli has been suggested to favorably enable representing emotions with less implications about awareness than with verbal expressions (Fischer et al., 2002) but some verbal guidance is recommended for using a SAM scale (Bradley & Lang, 1999a; Lang et al., 2008). With our online questionnaire we considered appropriate to provide guidance for using a SAM scale in a written format (see Lahti et al., 2017, Appendix B).

## 2.2 Sample and procedure

The current research was carried out in accordance with the recommendations of Finnish Advisory Board on Research Integrity and Helsinki University Hospital (HUS) Ethical Committee (permission number 93/13/03/03/2014) with written informed consent from all subjects in accordance with the Declaration of Helsinki. In our experimental evaluation with a group of 35 volunteer persons outside a clinical setting each person completed in November-December 2016 an online questionnaire which asked him/her to evaluate a total of 195 emotional adjectives and 16 pregnancy-related nouns with the Self Assessment Manikin (SAM) test. The questionnaire asked also the person to indicate the gender (man or woman), age, number of own children and native language. For each of the total of 211 concepts the person had to provide a rating value on a five-point visual scale for three affectivity dimensions of pleasure, arousal and dominance. The experiment was carried out in Finnish language and all participating persons had Finnish as a native language. Table 1 shows some background information about the group of 35 persons. There was no time limit for answering the SAM test. To evaluate all three dimensional properties the persons used on average 11.7 seconds for an emotional adjective and 12.0 seconds for a pregnancy-related noun.

INSERT TABLE 1 HERE.

The set of 195 emotional adjectives was created in the development project of a wellbeing start-up company Emotion Tracker lead by Camilla Tuominen with an aim to represent common emotional adjectives of everyday life. The set of 16 pregnancy-related nouns in Finnish was created by two authors of this article (LL & HT) with an aim to represent common pregnancy-related nouns. All 211 concepts are listed in the original order used in the rating task in Lahti et al. (2017, Appendix A) each with a unique coarse English translation. To avoid bias of polysemy for example the word "pregnancy" (in Finnish "raskaus") was on purpose preceded by the words "infant" and "fetus" to cause a priming effect so that the person could give the affective ratings for the semantic meaning "pregnancy" instead of "heaviness". Lahti et al. (2017, Appendix B) shows the visual SAM scale and main instructions provided for the person about the SAM test based on the instructions of Bradley and Lang (1994; 1999a). The dimensions of pleasure, arousal and dominance were represented in Finnish as "miellyttävyys", "kiihdyttävyys" and "hallinnan tunne" respectively.

## 2.3. Statistical methods

The data of affective rating answers of an emotional adjective for each person can be represented with a vector in a three-dimensional space that we refer to as an *emotional vector space*. The collected rating answers for each of three dimensions are represented as integer numbers in a range [-2, 2] in the order of rising affectivity and thus each of the three coordinates of the vector can have an integer value in a range [-2, 2]. When making analysis in respect to a population subgroup we have decided to form *an average vector of an emotional adjective* to represent an average opinion about the affective rating answers of the persons belonging to the subgroup for this emotional adjective in question. The average vector of an emotional adjective is computed as the average of vectors of this specific emotional adjective of the persons belonging to the subgroup.

To evaluate the properties and relationships of the affective rating answers we carried out various tests relying on statistical methods, including Pearson product-moment correlation coefficient measures, analysis of variance (ANOVA) tests, Euclidean distance measures, cosine similarity measures and the unsupervised algorithm of k-means clustering.

## 3. Results

### 3.1 Comparison of the current research data and other linguistic resources

For the full population of 35 persons Figure 1 shows the distribution of average vectors of emotional adjectives for all 195 emotional adjectives.

INSERT FIGURE 1 HERE.

We carried out a comparison of Pearson product-moment correlation coefficient measures for affective ratings between the pairs of the dimensions pleasure, arousal and dominance for several population subgroup categorizations in respect to 195 emotional adjectives and 16 pregnancy-related nouns (see Lahti et al., 2017, Appendix C, Table C1). A significant strong positive correlation ($r > 0.5$, $p < 0.001$) appeared between the dimensions of pleasure and dominance in respect to emotional adjectives and pregnancy-related nouns for the full population sample and for the subgroups women, men, "women with children" and "men without children" but not for the subgroup "women without children". In addition we evaluated the proportional distribution of 6825 rating answers of 35 persons in respect to 125 alternative answer categories based on three dimensions of affectivity for 195 emotional adjectives (see Lahti et al., 2017, Appendix C, Table C2).

We contrasted our current research data with other linguistic resources (see Lahti et al., 2017, Appendix A). In respect to 221 word lemmas having a frequency of at least two in Finnish emotion words of Tuovila (2005), 110 lemmas did not have a match with our 195 emotional adjectives. 440 Finnish nouns of Söderholm et al. (2013) had 55 matching lemmas with our 195 emotional adjectives and three matching lemmas with our 16 pregnancy-related nouns and 210 Finnish nouns of Eilola and Havelka (2010) had 35 matching lemmas with our 195 emotional adjectives and three matching lemmas with our 16 pregnancy-related nouns. We implemented comparisons between data sets with affective rating scales adjusted to a same range [-2,2].

Pearson product-moment correlation coefficient measures (see Lahti et al., 2017, Appendix C, Table C3) showed a significant strong positive correlation ($r > 0.5$, $p < 0.001$) for emotional adjectives between our current research data and the data of Söderholm et al. (2013) for the dimensions pleasure and arousal and between our current research data and the data of Eilola and Havelka (2010) for the dimension pleasure. However for pregnancy-related nouns a significant strong positive correlation ($r > 0.5$, $p < 0.001$) was not found between these data sets.

Pearson product-moment correlation coefficient measures (see Lahti et al., 2017, Appendix C, Table C4) showed a significant strong positive correlation ($r > 0.5$, $p < 0.05$) for the matching concepts in respect to the six basic emotion categories proposed by Ekman (1972) for the dimension pleasure when comparing our current research data with the data of Eilola and Havelka (2010) and with adjective and noun lemmas of Warriner et al. (2013) and for the dimension dominance when comparing our current research data with adjective and noun lemmas of Warriner et al. (2013).

We computed cosine similarity measures between average vectors of words of our current research data and Warriner et al. (2013) for the matching concepts in respect to the 16 pregnancy-related nouns (see Lahti et al., 2017, Appendix C, Table C8) and found that for ten of sixteen nouns the cosine similarity measure had a value over 0.5. Pearson product-moment correlation coefficient measures (see Lahti et al., 2017, Appendix C, Table C9) showed a significant strong positive correlation (r > 0.5, p < 0.05) for the matching concepts in respect to the 16 pregnancy-related nouns for the dimensions pleasure and dominance when comparing our current research data with the data of Warriner et al. (2013).

Besides the full population samples, we used partial segments of the full population samples to compare affective ratings of our current research data and the data of Warriner et al. (2013) especially for the words fetus ("sikiö"), pregnancy ("raskaus") and giving birth ("synnytys") that are essential in the context of maternal care (see Lahti et al., 2017, Appendix C, Table C10). A significant strong positive correlation (r > 0.5, p < 0.05) appeared only for the subgroup "women with children" and only for this same subgroup the cosine similarity measure reached a value over 0.5, in respect to the words pregnancy ("raskaus") and giving birth ("synnytys").

### 3.2 Categorization of affective ratings for subgroups

We analyzed the distributional variation of affective ratings for 195 emotional adjectives depending on alternative subgroup categorizations: gender, gender-parental (a combination of gender and having or not having child/children), rating-daytime (the time of the day when the person gave affective rating answers) and rating-duration (the average duration of giving three affective rating answers, computed based on the duration for all 195 emotional adjectives).

Our decision to use in our analysis these categorizations of the subgroups was motivated by our aim to identify such segmentations for affective ratings that can be used for emotional diary text analysis to support maternal care. Categorization of the subgroups based on the gender and having child/children enables observing population segments that we consider important to be addressed in maternal care. Categorization of the subgroups based on the time of the day and the average duration of giving three affective rating answers enables observing time-dependent behavior patterns related to reporting affectivity with a diary.

Concerning subgroups gender, gender-parental, rating-daytime and rating-duration for 195 emotional adjectives ANOVA tests indicated that belonging to a subgroup had same kinds of significant main effect patterns on affective rating dimensions for all these four subgroup categorizations (see Table 2 and Lahti et al., 2017, Appendix C). An ANOVA test indicated that the main effect of belonging to a subgroup on the combined set of all three affective rating dimensions was significant for the subgroups gender, $F(3, 6821) = 4.689$, $p < 0.003$, gender-parental, $F(3, 6626) = 5.8986$, $p < 0.001$, rating-daytime, $F(3, 6821) = 18.654$, $p < 0.001$, and rating-duration, $F(3, 6821) = 22.75$, $p < 0.001$.

For 195 emotional adjectives ANOVA tests indicated that the main effect of belonging to a gender subgroup on affectivity dimension was significant for arousal dimension, $F(1, 6823) = 7.8513$, $p < 0.006$, but non-significant for pleasure dimension, $F(1, 6823) = 2.9103$, $p > 0.05$, and dominance dimension, $F(1, 6823) = 3.7382$, $p > 0.05$. For 195 emotional adjectives ANOVA tests indicated that the main effect of belonging to a gender-parental subgroup on affectivity dimension was significant for arousal dimension, $F(1, 6628) = 10.94$, $p < 0.001$, and dominance dimension, $F(1, 6628) = 4.1632$, $p < 0.05$, but non-significant for pleasure dimension, $F(1, 6628) = 1.564$, $p > 0.05$. For the main effects of belonging to subgroups rating-daytime and rating-duration, see Lahti et al. (2017, Appendix C).

When aiming to replicate a previously noted pattern that typically men provide higher affective ratings than women and younger people higher ratings than older people in all three affectivity dimensions (Warriner et al. (2013), our ANOVA tests indicated that for 195 emotional adjectives the combined set of all three affective rating dimensions differed significantly in respect to gender subgroups women and men, $F(3, 2726) = 214.37$, $p < 0.001$, as well as gender-parental subgroups "women with children" and "women without children", $F(3, 1556) = 186.87$, $p < 0.001$, "women with children" and "men without children", $F(3, 1556) = 112.4$, $p < 0.001$, and "women without children" and "men without children", $F(3, 2531) = 198.01$, $p < 0.001$. For the significance of difference between gender subgroups and between gender-parental subgroups for each of three affectivity dimensions, see Lahti et al. (2017, Appendix C).

In an attempt to replicate at least indirectly the previous results of Rosebrock et al. (2015) we carried out ANOVA tests that indicated that the arousal ratings of the subgroup "women with children" and the subgroup "women without children" differed significantly for 195 emotional adjectives, $F(1, 1558) = 95.49$, $p < 0.001$, whereas they did not differ significantly for 16 pregnancy-related nouns, $F(1, 126) = 0.126$, $p > 0.05$ (see Lahti et al., 2017, Appendix C).

We aimed to replicate the previous results (Sereno et al., 2015; Unkelbach et al., 2008; Kuperman, 2015; Das et al., 2012) that in the case of high-frequency words the responses are faster for positive than negative or neutral words. Addressing the frequency effect (Duyck et al., 2008) we selected a subsample of ten positive and ten negative high-frequency words concerning 195 emotional adjectives based on Parole corpus (2017) and compared the response times for the affective rating of pleasure measured with the accuracy level of a second. Below a cut-off level of 10 seconds the response time in seconds for the set of positive words ($M = 2.95$, $SD = 1.46$) and for the set of negative words ($M = 3.02$, $SD = 1.61$) did not differ significantly, $F(1, 630) = 0.296$, $p > 0.05$ (see Lahti et al., 2017, Appendix C).

INSERT TABLE 2 HERE.

## 3.3 Evaluating general affective properties

### 3.3.1 Addressing varied subgroups

Since the results of ANOVA tests (chapter 3.2) indicated significant main effect and difference patterns concerning affective rating dimensions (pleasure, arousal and dominance) and belonging to alternative subgroup categorizations (in respect to subgroups of gender, gender-parental, rating-daytime and rating-duration) we thus decided to develop computational models that rely especially on these categorizations.

We describe next the formation of a general and a word-specific transformation model to define how the affective ratings used by a person belonging to a certain subgroup can be represented with affective ratings used by another person belonging to another subgroup (chapter 3.3). Then our further evaluation (chapters 3.4 and 3.5) suggests additional models to support interpreting affective ratings relying on comparison of extreme affective ratings and clusters of affective ratings so that we focus specifically on gender-parental subgroups. After that our formation of models extends from emotional adjectives to pregnancy-related nouns (chapter 3.6). Then finally we make some general remarks about emerging patterns of affective words that seem to be typical in respect to each proposed model when aiming to support maternal care (chapter 3.7).

### 3.3.2 General and word-specific transformation models

Based on the distinctive properties of affective rating answers of emotional adjectives in respect to alternative subgroups it is possible to formulate *a general transformation model* that aims to define how the affective ratings used by a person belonging to a certain subgroup (transmitting subgroup) can be represented with affective ratings used by another person belonging to another subgroup (receiving subgroup). The general transformation model can be practically defined as the summation of two vectors that are the average vector of an emotional adjective in respect to the transmitting subgroup $E_{transmitting}$ and *a general transformation vector* $T_{transmitting \rightarrow receiving}$.

This can be written as:

$E_{receiving}(w_i) = E_{transmitting}(w_i) + T_{transmitting \rightarrow receiving}$ .

The general transformation vector is thus the difference (i.e. subtraction) of the two average vectors of the emotional adjectives in question for the two different subgroups in the same emotional vector space. A general transformation vector in the opposite direction between a pair of subgroups can be gained by changing all coordinates into complement values.

For gender-parental subgroups one possible general transformation vector is $T_{women\_with\_children \rightarrow women\_without\_children} = (0; +0.15; +0.03)$.

As an extension of the general transformation model discussed above in respect to an average vector of an emotional adjective we now define *a word-specific transformation model* that enables to interpret the degree of correspondence of matching words between persons belonging to different subgroups. We define the word-specific transformation model as the summation of two vectors that are the average vector of a specific word in respect to the transmitting subgroup $WE_{transmitting}$ (word) and *a word-specific transformation vector* $WT_{transmitting \rightarrow receiving}$ (word) so that the result is the average vector of the matching word but now in respect to the receiving subgroup.

For gender-parental subgroups one possible word-specific transformation vector is:

$WT_{women\_without\_children \rightarrow women\_with\_children}$ ("anxious") = (+0.35; -0.85; +0.06)

More general and word-specific transformation vectors are shown in Lahti et al. (2017, Appendix D).

### 3.4 Evaluating extreme affective properties

### 3.4.1 Extreme corners of the emotional vector space

By computing the Euclidean distances between the average vectors we generated for each gender-parental subgroup lists of average vectors of emotional adjectives that have a direction pointing into one of the eight corners of the three-dimensional emotional vector space (see Lahti et al., 2017, Appendix E).

When observing the eight corners of the emotional space with values more or less than zero for each dimension (pleasure, arousal and dominance) it appears that a majority of average vectors for 195 emotional adjectives is positioned in the corners that make a requirement "dominance < 0 and pleasure < 0" or "dominance > 0 and pleasure > 0". When observing the eight corners of the emotional space with values more than 1 or less than -1 it appears that only less than 25 of 195 emotional adjectives have an average vector pointing to an extreme corner of the emotional vector space.

Table 3 lists some distinctive average vectors of emotional adjectives towards the eight corners of the emotional vector space with affective rating values more or less than zero in six pair-wise comparison combinations between three gender-parental subgroups. Table 3 reports for each gender-parental subgroup such average vectors of emotional adjectives that did emerge in the current corner of the emotional vector space for this subgroup but did not for the compared subgroup, listed here in the decreasing order of the distance between the average vectors of this emotional adjective for both subgroups now in comparison (the three adjectives with the highest distance are shown, if available for this pair of subgroups).

For example a comparison shows that in the corner of emotional vector space that makes a requirement "pleasure < 0, arousal < 0 and dominance < 0" there is an average vector of the word "uncomfortable" for the subgroup "Women with children (wwc)" but not for the subgroup "Women without children (wwoc)". We suggest that this finding can indicate that women with children may on average interpret the affectivity associated with the word "uncomfortable" as a more distinctive and typical for this specific corner of the emotional vector space (i.e. for this type of combination of values of pleasure, arousal and dominance) than women without children interpret.

Similarly Lahti et al. (2017, Appendix E) shows some distinctive average vectors of emotional adjectives towards the eight corners of the emotional vector space that make a requirement of values over 1 or less than -1 for each dimension (pleasure, arousal and dominance).

INSERT TABLE 3 HERE.

### 3.4.2 Extreme affective shift model

Based on the results above it is possible to formulate an *extreme affective shift model* that aims to define how in the extreme corners of the emotional vector space this model can represent the strongest shifts of average vectors of emotional adjectives between the gender-parental subgroups.

We define the extreme affective shift model by a list of highest-ranking distinctive average vectors of emotional adjectives towards a corner of the emotional vector space for a compared pair of gender-parental subgroups, for example written in the case discussed above (in chapter 3.4.1) as:

$L$(pleasure < 0 & arousal < 0 & dominance < 0; wwc & ¬wwoc) = { "uncomfortable" (1.18); "shared sense of shame" (1.1); "awkward" (1.06) } .

More extreme affective shift models are shown in Lahti et al. (2017, Appendix E).

### 3.5. Evaluating affective properties of clusters

### 3.5.1 K-means clustering of the current research data

By computing k-means clustering for the gathered affective rating data we wanted to investigate how the data could be automatically categorized for three gender-parental subgroups and how this categorization could enable developing ways to support reliable communication in medical context about affective expressions between persons belonging to different subgroups.

The unsupervised algorithm of k-means clustering is a popular computational method to make partitioning of observations into separate clusters relying on the principle that each observation belongs to the cluster having the nearest mean value (Lloyd, 1957). The k-means clustering can be considered to represent Gaussian mixture modeling, expectation-maximization, minimizing variance and dividing the data space into Vornoi cells. Performing the k-means clustering aims to minimize the pooled within-cluster sum of squares around the cluster means thus minimizing the squared Euclidean distance.

We generated the gap statistic (Tibshirani et al., 2001; R Project, 2017; Factoextra library, 2017) for our affective rating data which peaked clearly at the value 5 for each of three gender-parental subgroups thus suggesting that value 5 can be an optimal number of clusters. Then we generated k-means clustering for our affective rating data with a condition to get exactly five clusters for each gender-parental subgroup (Cluster library, 2017). Our initial experiments motivated us to pre-process our data before clustering by centering (the column means were subtracted from their corresponding columns) and scaling (the centered columns were divided by their standard deviations) (R script scale method, 2017). Although the clustering algorithm produced for affective rating data new coordinates positioned in a centered and scaled coordinate system we report our clustering results by referring to affective rating data expressed in the original non-centered and non-scaled coordinate system of the emotional vector space.

Figure 2 shows the generated five clusters of affective rating data for each gender-parental subgroup as a two-dimensional visualization that makes a projection from three-dimensional space to two-dimensional space based on principal component analysis (Factoextra library, 2017).

INSERT FIGURE 2 HERE.

By a pair-wise comparison of data sets of clusters we identified the five corresponding most matching clusters between gender-parental subgroups that we refer to with notations A, B, C, D and E, supplied with a subindex to indicate the subgroup currently in question, for example $A_{\text{Women\_with\_children}}$. We computed for all five clusters of each gender-parental subgroup the measures of precision, recall and F1 to indicate the amount of matching of clustering of subgroups, as shown in Lahti et al. (2017, Appendix C).

For each gender-parental subgroup we defined *an average vector of a cluster* to represent average opinion about the affective rating answers of the emotional adjectives belonging to this specific cluster of the subgroup in question (see Table 4). The average vector of a cluster is computed as the average of vectors of the emotional adjectives belonging to this specific cluster of the subgroup in question.

INSERT TABLE 4 HERE.

### 3.5.2 Cluster-based transformation model

As an extension of the general transformation model discussed above in respect to an average vector of an emotional adjective we now define *a cluster-based transformation model* that enables to interpret the degree of correspondence of matching clusters between different gender-parental subgroups. We define the cluster-based transformation model as the summation of two vectors that are the average vector of a cluster in respect to the transmitting subgroup $CE_{\text{transmitting}}$(cluster name) and *a cluster-based transformation vector* $CT_{\text{transmitting}\rightarrow\text{receiving}}$(cluster name) so that the result is the average vector of the matching cluster but now in respect to the receiving subgroup.

For example for gender-parental subgroups one possible cluster-based transformation vector is:

$CT_{\text{women\_without\_children}\rightarrow\text{women\_with\_children}}$ (cluster A) = (+0.074; -0.169; +0.129)

More cluster-based transformation vectors are shown in Lahti et al. (2017, Appendix D).

### 3.5.3 Scales of average vectors of clusters

For each of five clusters of each gender-parental subgroup we identified a ranking order list about all the emotional adjectives belonging to this specific cluster ordered in respect to an increasing distance of the average vector of this emotional adjective from the average vector of the cluster in question (see Table 5, three closest average vectors of emotional adjectives are shown).

Besides using the distance between two vectors it is possible to use also the directional angle between them to form ranking order lists about emotional adjectives. For each cluster of a gender-parental subgroup we identified those emotional adjectives that have exactly or nearly the same direction as the average vector of the cluster (based on a threshold of a cosine similarity value of at least 0.99) listed in an ascending order in respect to the distance of the average vector of the emotional adjective from the origo (see Lahti et al., 2017, Appendix F). In these lists we indicated with a prefix > three emotional adjectives having the shortest distance to the average vector of a cluster (listed also in Table 5 for the cluster in question) and with a suffix @ emotional adjectives closest to the average vector of a cluster even if they are not having a cosine similarity value of at least 0.99.

INSERT TABLE 5 HERE.

We suggest that each of these ordered lists of emotional adjectives can be considered to define a scale of continuity of emotional expressions that are associated with each other and that certain fundamental properties about this association may manifest well in the affective properties of the average vector of the cluster in question. Furthermore we suggest that the emotional adjectives belonging to the same scale can be considered to manifest the degree of strength of the affective properties of the average vector of the cluster in question so that for shorter average vectors of emotional adjectives the strength of these affective properties is relatively low and for longer average vectors of emotional adjectives the strength of these affective properties is relatively high.

In addition it is possible to consider also such average vectors of emotional adjectives that have exactly or nearly the opposite direction as the average vector of the cluster and we suggest that these emotional adjectives manifest an especially low strength of the affective properties of the average vector of the cluster in question or even manifest opposite affective properties.

### 3.5.4 Scale-cluster model

As a variation of the extreme affective shift model discussed above in respect to identify some essential rating differences in extreme affective expressions between subgroups we now define *a scale-cluster model* that enables to interpret the degree of correspondence of average vectors of emotional adjectives relying on a shared consecutive scale from the origo towards an average vector of a cluster for a gender-parental subgroup.

We define the scale-cluster model by lists SL(subgroup; cluster name) of average vectors of emotional adjectives from the origo towards an average vector of a cluster for a gender-parental subgroup in question (supplied with the values of distance from the origo), for example written in the case shown in Lahti et al. (2017, Appendix F) as:

SL (women without children; cluster A) = { "> admiring @" (1.41); "> curious" (1.46); "> gritty" (1.72); "hopeful" (1.72); "amazing" (1.8); "superb" (1.93); "fantastic" (2.07) } .

More scale-cluster models are shown in Lahti et al. (2017, Appendix F).

### 3.6 Average vectors of pregnancy-related nouns

### 3.6.1 Pregnancy-related nouns for gender-parental subgroups

In a similar way as done above for the emotional adjectives we formed an average vector for each pregnancy-related noun to represent average opinion about the affective rating answers of the persons belonging to the subgroup for this pregnancy-related noun in question. The average vector of a pregnancy-related noun is computed for each gender-parental subgroup as the average of vectors of this specific pregnancy-related noun of the persons belonging to the subgroup. Table 6 shows for each gender-parental subgroup the 16 average vectors of a pregnancy-related noun and the nearest average vectors of clusters in an ascending order in respect to the distance between the average vector of the pregnancy-related noun and the average vector of a cluster.

The word-specific transformation model already discussed above (chapter 3.3.3) enables to interpret the degree of correspondence of matching words between different gender-parental subgroups also in respect to pregnancy-related nouns, for example one possible word-specific transformation vector is:

$WT_{women\_without\_children \rightarrow women\_with\_children}$ ("intimate relationship") = (-0.18; +0.33; +0.59)

More word-specific transformation vectors are shown in Lahti et al. (2017, Appendix D).

INSERT TABLE 6 HERE.

### 3.6.2 Attraction-cluster model

As a variation of the scale-cluster model discussed above in respect to the average vectors of emotional adjectives relying on a shared consecutive scale we now define *an attraction-cluster model* that enables to interpret the relatedness of an average vector of a word to the nearest average vectors of clusters for a gender-parental subgroup.

We define the attraction-cluster model by lists AL(subgroup; word) of the nearest average vectors of clusters for the gender-parental subgroup in question in respect to a word in the order of ascending

distance, for example written in the case shown in Table 6 as:

AL (women without children; "intimate relationship") = { C, A, D, B, E } .

More lists of attraction-cluster models are shown in Lahti et al. (2017, Appendix F).

### 3.7 Emerging patterns of affectivity in respect to the proposed models

In respect to all models we proposed in chapters 3.3-3.6 we can make summarizing remarks about some of the highest-ranking subgroups, words and clusters concerning emerging patterns of affectivity for each model (see Table 7). These emerging patterns of affectivity can be contrasted with our aim to support maternal care and emphasize gender-parental subgroups "women without children", "women with children" and "men without children".

INSERT TABLE 7 HERE.

### 4. Discussion

### 4.1 Main findings

To enable identification of distinctive emotional patterns in language usage of population segments there is a need to gather a reference collection of affective ratings for the population segments in question. Outside a clinical setting we have experimentally gathered affective ratings for 195 emotional adjectives and 16 pregnancy-related nouns in Finnish language ($n_{persons\_giving\_ratings}$=35) in dimensions of pleasure, arousal and dominance.

In respect to the previous research evaluating the degree of dependence between the three affectivity dimensions (pleasure, arousal and dominance) (Mehrabian, 1972; Bradley & Lang, 1994; Bradley & Lang, 1999a; Warriner et al., 2013; Montefinese et al., 2014; Moors et al., 2013; Imbir, 2015), our affective rating data indicated a significant strong correlation between the dimensions of pleasure and dominance in respect to emotional adjectives and pregnancy-related nouns for the full population sample and for the subgroups women, men, "women with children" and "men without children" but not for the subgroup "women without children" (see Lahti et al., 2017, Appendix C, Table C1).

For 195 emotional adjectives ANOVA tests indicated that the main effect of belonging to a subgroup on the combined set of all three affective rating dimensions was significant for the subgroups gender, gender-parental, rating-daytime and rating-duration (see Lahti et al., 2017, Appendix C). Further ANOVA tests indicated various main effects of belonging to a subgroup on affective rating dimensions significantly as reported in chapter 3.2.

Our ANOVA tests indicated that for 195 emotional adjectives the combined set of all three affective rating dimensions differed significantly in respect to gender subgroups women and men as well as gender-parental subgroups "women with children" and "women without children", "women with children" and "men without children", and "women without children" and "men without children". Further ANOVA tests indicated various significant differences of affectivity dimensions between gender subgroups and between gender-parental subgroups as reported in chapter 3.2. Our ANOVA tests indicated that the arousal ratings of the gender-parental subgroups "women with children" and "women without children" did differ significantly for 195 emotional adjectives whereas they did not differ significantly for 16 pregnancy-related nouns.

Therefore in respect to the pattern that men provide higher affective ratings than women and younger people higher ratings than older people in all three affectivity dimensions (Warriner et al. 2013) and that the pregnant and postpartum women and the age-matched women expressed lower arousal ratings relative to the college-aged women (Rosebrock et al. 2015), our results indicate accordance with significant affectivity differences of gender, age and motherhood.

When aiming to replicate with our affective rating data the previous results that in the case of high-frequency words the responses are faster for positive than negative words (Sereno et al., 2015; Unkelbach et al., 2008; Kuperman, 2015; Das et al., 2012) the response times for the set of positive words and for the set of negative words did not differ significantly (see Lahti et al., 2017, Appendix C).

In the context of Finnish language usage we contrasted our current research data with the collection of Finnish emotion words of Tuovila (2005) and affective rating collections of Söderholm et al. (2013) and Eilola and Havelka (2010) thus identifying some clear lack of overlap in coverage of corresponding vocabularies and varied use of conjugated forms of lemmas and synonyms (see Lahti et al., 2017, Appendix A). This result highlighted the challenge of balancing with sufficiently compact but also extensive collections of linguistic data and to be able with reasonable computational resources to identify useful linguistic patterns when aiming to analyze large data sets of health-related communication. Furthermore informal language usage containing for example slang or dialects can pose additional challenges for modeling language usage patterns and thus requires innovative solutions.

When contrasting with previously identified protruding patterns in categorization of emotion words our affective rating data did not follow clearly all these patterns in respect to results of emotional vector space for example based on distance, direction or clustering. In the collection of Finnish emotion words (Tuovila, 2005) the six highest-frequent emotional Finnish lemmas in a decreasing order were joy ("ilo"; 4.2 % of response items), love ("rakkaus"; 4.1 %), anger ("viha"; 3.8 %), sorrow ("suru"; 3.6 %), fear ("pelko"; 2.5 %) and happiness ("onnellisuus"; 2.1 %) thus four of them matching with the six basic emotion categories proposed by Ekman (1972) but love ("rakkaus") and happiness ("onnellisuus") surpassed disgust ("inho"; 1.1 %) and surprise ("hämmästys"; 0.4 %; or "yllättyneisyys"; 0.1 %) (Tuovila, 2005).

As reported in chapter 3.1, we carried out a comparison of Pearson product-moment correlation coefficient measures for affective ratings between emotional adjectives of our current research data and the corresponding emotional words in data collections of Söderholm et al. (2013), Eilola and Havelka (2010) and Warriner et al. (2013) as well as in respect to the six basic emotion categories proposed by Ekman (1972) (see Lahti et al., 2017, Appendix C). We found a significant strong positive correlation for emotional adjectives between our current research data and the data of Söderholm et al. (2013) for the dimensions pleasure and arousal and between our current research data and the data of Eilola and Havelka (2010) for the dimension pleasure (see Lahti et al., 2017, Appendix C, Table C3). However we did not find for pregnancy-related nouns a significant strong correlation between these data sets (see Lahti et al., 2017, Appendix C, Table C3).

We found a significant strong positive correlation for matching concepts in respect to the six basic emotion categories proposed by Ekman (1972) for the dimension pleasure when comparing our current research data with the data of Eilola and Havelka (2010) and with adjective and noun lemmas of Warriner et al. (2013) and for the dimension dominance when comparing our current research data with adjective and noun lemmas of Warriner et al. (2013) (see Lahti et al., 2017, Appendix C, Table C4).

Comparison of our current research data and the data of Warriner et al. (2013) for the 16 pregnancy-related nouns showed a significant strong positive correlation for the dimensions pleasure and dominance (see Lahti et al., 2017, Appendix C, Table C9) and that for ten of sixteen nouns the cosine similarity measure had a value over 0.5 (see Lahti et al., 2017, Appendix C, Table C8). In a further analysis with partial segments of the full population samples for words fetus ("sikiö"), pregnancy ("raskaus") and giving birth ("synnytys"), a significant strong positive correlation appeared only for the subgroup "women with children" and only for this same subgroup the cosine similarity measure reached a value over 0.5, in respect to the words pregnancy ("raskaus") and giving birth ("synnytys")

Our data analysis has identified how affective ratings depend on various properties of language usage and population subgroup categorization, including gender-parental subgroups "women without children", "women with children" and "men without children" that we consider important population segments to be addressed in maternal care. Based on distributional variation of affective ratings in respect to subgroups we developed a set of models to support interpretation and communication of affective expressions between persons who experience and characterize affectivity of concepts differently (see chapters 3.3-3.6).

We suggest that to convey three-dimensional affective ratings in the emotional vector space from one person to another it is practical to have a general transformation model that can convert coordinates for the whole affective rating data set at once. Besides the general transformation model we also suggest using the word-specific transformation model to address distinctive individual patterns of correspondence in conversion of affective rating coordinates. The transformation vectors defined for each transformation between affective representations of subgroups enable for two persons belonging to different subgroups to have a shared scale for expressing and interpreting reliably affective linguistic expressions when they communicate. Thus in respect to research of interaction in healthcare the transformation vectors enable for example a patient and a medical professional who belong to different subgroups to have a mapping between their affective expressions concerning the state of health and thus to ensure that needs of the patient can be appropriately addressed.

However since the relation between the corresponding distributions of affective ratings is often non-uniform we suggest an additional coordinate conversion adjustment with the extreme affective shift model. It is possible that some of the serious and unexpected misunderstandings can happen with extreme affective ratings interpreted distinctively differently. The lists of the extreme affective shift model can be computationally exploited to generate support for reliable communication of affective linguistic expressions. Based on the lists it is possible to identify some essential rating differences in extreme affective expressions between persons belonging to different subgroups and thus in medical context to avoid misinterpretation of affective expressions concerning the state of health.

Furthermore we suggest that an important way to support conversion of non-uniform coordinate distributions of affective ratings between persons can exploit a mapping between clusterings of affective expressions as is done with the cluster-based transformation model.

Based on the clustering scheme of affective ratings we developed a scale-cluster model that enables to interpret the degree of correspondence of affective ratings relying on a shared consecutive scale with coordinates ranging from the origo towards a cluster center. The lists of scale-cluster model can be computationally exploited to generate support for reliable communication of affective linguistic expressions. Based on the lists it is possible to identify some essential relatedness and degree of rising strength of ratings in affective expressions between persons belonging to different subgroups and thus in medical context to avoid misinterpretation of affective expressions concerning the state of health.

In addition the attraction-cluster model was developed to enable interpreting the relatedness of affective rating of a concept to the nearest cluster centers of affective ratings. The lists of attraction-cluster model can be computationally exploited to generate support for reliable communication of affective linguistic expressions. Based on the lists it is possible to identify some essential relatedness and clustering of ratings in affective expressions between persons belonging to different subgroups and thus in medical context to avoid misinterpretation of affective expressions concerning the state of health.

Our affective rating data shows various patterns of ratings of affective expressions that depend on many characteristics of persons or subgroups in question. These findings can be contrasted with the results of previous research concerning diverse language usage patterns identified in health-related communication which indicate how essential challenge it is to develop methods that can support avoiding misunderstanding in communication.

In the analysis of health-related communication based on NHS Direct Corpus it was identified that protruding keywords after filtering out medical jargon can be classified into conceptual categories of negatives, imperatives, pronouns, vague language, affirmations (positive backchannels) and directives, and furthermore the occurrences of keywords seemed to have specific patterns emerging in certain phases of consultation, such as assessment, advice giving and wrapping up the conversation (Adolph et al., 2004). In the analysis of health-related communication based on Teenage Health Freak corpus it was identified that protruding keywords in health-related inquiries from teenagers were about communication, and even if the terms did not directly imply negative or positive evaluation the collocation terms for them suggested negative viewpoint and having concerns about confidentiality, disclosing problems and embarrassment (Harvey et al., 2008).

When contrasting two health advice websites Teenage Health Freak and Lucy Anwers in respect to reproductive organs health in a subtopic of HIV/AIDS it was found that the distribution of relative frequencies of question types and used terminology varied considerably (Harvey et al., 2013). For example the question messages had an average length of 16 words for Teenage Health Freak and 159 words for Lucy Anwers (Harvey et al., 2013).

Our affective rating data also shows how patterns and differences of affectivity can be hard to identify reliably with a relatively limited data set thus making also hard to make segmentation of patients based on the affectivity. This result can be contrasted with the previous notion that the relative sparseness of health-related data can make it challenging to extract accurate knowledge about a person's health (Paul & Dredze, 2012; Parker, Wei, Yates, Frieder & Goharian, 2013; Jiang et al., 2016). When a set of over 2 billion Twitter messages was matched with a list of 20 000 key phrases related to illnesses/diseases it was possible to gain a set of 11.7 million messages of which 24.4 % indicated that the user had an acute illness and 11.3 % had a general comment about the health of the user or someone else (Paul & Dredze, 2012). Thus it can be approximated that in public online communication self-reporting about personal health may form about 0.14 % of average messages.

Even if we recognize the specific need to address culturally dependent characteristics emerging in Finnish language we identified positive accordance between our affective rating collection of 211 Finnish words and the collection of Warriner et al. (2013) having an impressive coverage of 13 915 English language lemmas. However this comparison also showed that even when matching semantically fine-nuanced corpora it is possible that various culturally confusing discrepancies or distortions of interpreting affectivity can emerge when using word-based affective ratings especially if many influencing background conditions of data acquisition of affective ratings remain uncontrollable or not documented.

Thus it appears that there is a need to develop solutions that can address affective ratings reliably also for broader ranges of linguistic items than individual words (including phrases, addressing homonymy and varied usage contexts) and furthermore to develop reliable dynamic affective ratings for on-demand use for specifically selected persons and tailored usage contexts. These findings seem to indicate that to develop fine-tuned models to support maternal care based on analyzing self-reported emotional diary texts of pregnant women there is a strong need to gather extensive collections of affective rating data directly from pregnant women attending maternal care.

Our gained results related to the development of methods for mobile measurement of affectivity in maternal care provide a diverse collection of linguistic affectivity data. Our process of collecting this data was motivated by previously suggested responsiveness of pregnant women in respect to web-based applications (Cormick et al., 2012; Parker, Dmitrieva, Frolov & Gazmararian, 2012) and ecological momentary assessments (Bolger et al., 2003). Furthermore the behavioral properties of our collected data can be considered to reflect the principles identified in the previous research concerning the constructed emotion theory suggesting that the brain uses previous experiences of an emotion to categorize sensory input to construct an instance of emotion (Barrett et al., 2015; Barrett, 2017).

Our collected data contains various relatively complex personal and behavioral patterns that can possibly indicate similarly challenging needs for analysis as was identified in the previous research concerning analyzing informal patient-driven self-reporting and health-related communication (Adolph et al., 2004; Harvey et al., 2013) and analyzing unconstrained language of personal health-related communication (Paul & Dredze, 2012; Parker, Wei, Yates, Frieder & Goharian, 2013; Jiang et al., 2016).

Our findings and their comparison to previous research indicate a need to develop further methods to analyze in a complementing way the affective ratings of the patient, the present clinical data and earlier patient records data to discover how varied events along pregnancy both in everyday life and associated with clinical visits can be identified and supported based on distinctive patterns emerging in affective expressions.

## 4.2 Practical implications

We suggest that by identifying the emotional state it is possible to access useful information concerning the patient's motivation and orientation towards his/her life and healthcare services, including satisfaction, intensity and the feel of having a sufficient control. Temporal progressive analysis of the patient's emotional states and contrasting them with other background information enables creating predictive models about the patient's health and its relation to measured emotional states. Maternal care is a domain of life highly associated with concerns of the future and affectively sensitive events that can benefit from providing personal support to address needs indicated by the emotional states. However to accomplish a reliable acquisition of affective ratings as well as interpretation and communication of them between persons in the maternal care context there is a need for implementing and validation of appropriate computational models and methods. In this article we have reported an experimental evaluation of data acquisition and analysis methods that enabled us to develop new models to support measuring, interpretation and communication of affective expressions between population subgroups typically present in maternal care.

Our proposed models should be preferably considered as modular components that can be flexibly modified and combined to enable new forms of support for reliable communication between persons having different affective rating systems. Thus our proposed models aim to address the principles of constructed emotion theory (Barrett et al., 2015; Barrett, 2017) by recognizing that there are unlimited ways to interpret emotions by persons and that various properties of the past and current behavior of persons influence how they experience emotions.

We have developed our models to enable finding favorable ways to identify and support pregnant women that may have health problems such as suffering from mood disorders, psychosocial stress and antenatal anxiety. Motivated by the previous research (Lee et al., 2016; Bos et al., 2013; Nie et al., 2017; Rosebrock et al., 2015; Edwards et al., 2017; Lever Taylor et al., 2016; Posner et al., 2005; Feldman, 1995) and our reported results we suggest that our models can be exploited for developing computer-assisted methods to help both the pregnant woman and other persons participating in the maternal care process to understand better the properties of affective experiences and their dependencies to medical risks and conditions as well as to manage beneficially affectivity with emotion regulation strategies.

## 4.3 Limitations and future work

Even if the sample sizes of the subgroups of our research remain relatively low it was considered that this extent of accumulating observational data can be sufficient for the current research setting that aims to formulate further more extensive evaluation based on the initial findings gained from the current experiment.

In respect to affective ratings it needs to be noted that for gender-parental subgroups some identified distinctive properties that we have reported may not necessarily depend unambiguously on the gender or having or not having children. One reason for this uncertainty is that the age distributions are not fully balanced and previous research has reported that younger people provide higher affective ratings than older people in all three affectivity dimensions (Warriner et al., 2013).

Based on our current promising experimental results we suggest that the used and developed methods and models can be experimentally tested next in a real clinical setting with possibilities to develop further models specific to supporting affective language usage in maternal care.

Besides using our set of new developed models to support interpretation and communication of affective expressions between persons having different affective rating systems an other important way to exploit the identified distinctive differences in affective ratings between subgroups is in the domain of screening to enable segmenting subgroups of the population. In the context of maternal care by analyzing self-reported emotional diary texts of pregnant women it can be possible to identify specific patient segments based on the distinctive patterns of affective ratings in the person's language usage relying on tailored models developed in a similar way as we have developed our current models. This kind of automated analysis of self-reported emotional diary texts can help early detection of symptoms and risks of patients and to provide needed care efficiently for them.

## 5. Conclusion

Advancing technology opens increasing possibilities for analyzing health-related data collected by the patients themselves in everyday life with mobile devices. Therefore we have carried out an experimental evaluation outside a clinical setting to develop new computational models to support maternal care so that reliable interpretation and communication of affective expressions is possible between persons who represent different affective rating systems.

On a practical level, a conclusion based on our research is that belonging to a specific subgroup of persons has an effect on the type of emotions (affectivity) a person associates with certain concepts but this effect may appear only for some subgroups and some concepts with varying levels of strength. Thus we identified that persons belonging to different subgroups associate differently the affectivity measures for a set of concepts and we have reported what kind of patterns and dependencies can be found for these differences.

Our experimentally gathered affective ratings showed accordance with some previous research results (for example pleasure-dominance correlation and higher affective ratings for men and younger persons) but on the other hand comparison of our gathered affective ratings with collections of Finnish emotion words and affective ratings indicated some lack of overlap in coverage and confusing discrepancies.

Therefore to develop fine-tuned models to support maternal care it is recommended for future work to gather extensive collections of affective rating data directly from pregnant women attending maternal care and to analyze affective ratings in real clinical context to identify distinctive patterns in affective expressions indicating needs to be addressed.

## Conflict of interest

The authors declare that the research was conducted in the absence of any commercial or financial relationships that could be construed as a potential conflict of interest.

## Author contributions

## Funding

The current research was financially supported by Aalto University, Helsinki University Hospital and Tekes G3 program in Finland.

## Acknowledgements

The authors want to express gratitude to Camilla Tuominen for an essential role in creating and enabling free use of the list of 195 emotional adjectives.

## References

Adolphs S, Brown B, Carter R, Crawford P, Sahota O. (2004). Applied clinical linguistics: corpus linguistics in health care settings. *J Appl Linguist* 2004; 1:9–28. http://www.brown.uk.com/publications/adolphs.pdf

Azevedo, M., Silva, C., Dias, M. (1991). "Perfil de Estados de Humor": Adaptacao a Populacao Portuguesa. Psiq Clin. 1991; 12: 187-93.

Barrett, L. F. (2017). The theory of constructed emotion: An active inference account of interoception and categorization. *Social Cognitive and Affective Neuroscience*, 2017, 1–23. doi: 10.1093/scan/nsw154. https://academic.oup.com/scan/article/doi/10.1093/scan/nsw154/2823712/The-theory-of-constructed-emotion-an-active

Barrett, L. F., Wilson-Mendenhall, C. D., & Barsalou, L. W. (2015). The conceptual act theory: A road map. In L. F. Barrett & J. A. Russell (Eds.), The Psychological Construction of Emotion. New York: Guilford.

Bastos, A. M., Usrey, W. M., Adams, R. A., Mangun, G. R., Fries, P., & Friston, K. J. (2012). Canonical microcircuits for predictive coding. *Neuron*, 76(4), 695-711.

Boiger, M., & Mesquita, B. (2012). The construction of emotion in interactions, relationships, and cultures. *Emotion Review*, 4.

Bolger, N., Davis, A., & Rafaeli, E. (2003). Diary methods: Capturing life as it is lived. *Annual Review of Psychology*, 54, 579–616.

Bradley, M.M., & Lang, P.J. (1994). Measuring emotion: The self-assessment Manikin and the semantic differential. *J. Behav. Ther. & Exp. Psychiat.* Vol. 25, No. I. pp. 49-59.

Bradley, M.M., & Lang, P.J. (1999a). Affective norms for English words (ANEW): Instruction manual and affective ratings. Technical Report C-1, The Center for Research in Psychophysiology, University of Florida.

Bradley, M. M., & Lang, P. J. (1999b). International affective digitized sounds (IADS): Stimuli, instruction manual and affective ratings (Tech. Rep. No. B-2). Gainesville, FL: The Center for Research in Psychophysiology, University of Florida.

Bradley, M. M., & Lang, P. J. (2007). Affective Norms for English Text (ANET): Affective ratings of text and instruction manual. (Tech. Rep. No. D-1). University of Florida, Gainesville, FL.

Bressler, S. L., & Richter, C. G. (2015). Interareal oscillatory synchronization in top-down neocortical processing. *Current Opinion in Neurobiology*, 31, 62-66.

Candlin, C.N. & Candlin, S. (2003) Health care communication: a problematic site for applied linguistics research. *Annual Review of Applied Linguistics* 23: 134-154.

Clark, A. (2013). Whatever next? Predictive brains, situated agents, and the future of cognitive science. *Behavioral and Brain Sciences*, 36, 281-253.

Cluster library (2017). R language library "cluster". https://cran.r-project.org/web/packages/cluster/cluster.pdf

Cormick, G., Kim, N. A., Rodgers, A., Gibbons, L., Buekens, P. M., Belizán, J. M., (2012). Interest of pregnant women in the use of SMS (short message service) text messages for the improvement of perinatal and postnatal care. *Reproductive Health*, 6, 9.

Das, E., Vonkeman, C., and Hartmann, T. (2012). Mood as a resource in dealing with health recommendations: how mood affects information processing and acceptance of quit-smoking messages. *Psychol. Health* 27, 116–127. doi: 10.1080/08870446.2011.569888

Deneve, S. (2008). Bayesian Spiking Neurons I: Inference. *Neural Comput*, 20, 91-117.

Deneve, S., & Jardri, R. (2016). Circular inference: mistaken belief, misplaced trust. *Current Opinion in Behavioral Sciences*, 11, 40-48.

Drew, P., Chatwin, J. & Collins, S. (2001) Conversation analysis: A method for research into interactions between patients and health care professionals, *Health Expectations*, 4, 58-70.

Duyck, W., Vaderelst, D., Desmet, T., & Hartsuiker, R. (2008). The frequency effect in second-language visual word recognition. *Psychonomic Bulletin & Review*, 15(4), 850-855. http://users.ugent.be/~wduyck/articles/DuyckVanderelstDesmetHartsuiker2008.pdf

Edwards ES, Holzman JB, Burt NM, Rutherford HJV, Mayes LC, Bridgett DJ (2017). Maternal Emotion Regulation Strategies, Internalizing Problems and Infant Negative Affect. J Appl Dev Psychol. 2017 Jan-Feb;48:59-68. doi: 10.1016/j.appdev.2016.12.001. Epub 2016 Dec 29.

Eilola T. M., Havelka J. (2010). Affective norms for 210 British English and Finnish nouns. *Behav. Res. Methods* 42 134–140. 10.3758/BRM.42.1.134 (http://link.springer.com/article/10.3758/BRM.42.1.134)

Ekman, P. (1972). Universals and cultural differences in facial expressions of emotion. In J. Cole (Ed.), Nebraska Symposium on Motivation, 1971 (Vol. 19, pp. 207–283). Lincoln: University of Nebraska Press.

Elwyn, G. (2001) Shared Decision Making: Patient Involvement in Clinical Practice. Nijmegen: WOK.

Factoextra library (2017). R language library "factoextra". https://cran.r-project.org/web/packages/factoextra/factoextra.pdf

Feldman, LA. (1995). Valence focus and arousal focus: Individual differences in the structure of affective experience. Journal of Personality and Social Psychology, 69, 153–166.

Fischer, L., Brauns, D., & Belschak, F. (2002). Zur Messung von Emotionen in der angewandten Forschung Analysen mit den SAMs. Lengerich: Pabst-Science-Publishers.

Gallivan, J. P., Logan, L., Wolpert, D. M., & Flanagan, J. R. (2016). Parallel specification of competing sensorimotor control policies for alternative action options. *Nat Neurosci*, 19(2), 320-326. doi:10.1038/nn.4214

Harvey K1, Churchill D, Crawford P, Brown B, Mullany L, Macfarlane A, McPherson A. (2008). Health communication and adolescents: what do their emails tell us? *Fam Pract.* 2008 Aug;25(4):304-11. doi: 10.1093/fampra/cmn029. Epub 2008 Jun 17.

Harvey, K., Locher, M., & Mullany, L. (2013). "Can I Be at Risk of Getting AIDS?" A Linguistic Analysis of Two Internet Columns on Sexual Health. *Linguistik online* 59, 2/2013. http://www.linguistik-online.com/59_13/harveyLocherMullany.html

Hassanpour, S., & Langlotz, C. P. (2016). Unsupervised Topic Modeling in a Large Free Text Radiology Report Repository. *Journal of Digital Imaging*, 29(1), 59–62. http://doi.org/10.1007/s10278-015-9823-3. https://www.ncbi.nlm.nih.gov/pmc/articles/PMC4722022/

Hodes, R.L., Cook, E.W., & Lang, P.J. (1985). Individual differences in autonomic response: Conditioned association or conditioned fear? *Psychophysiology*, 22, 545-560.

Huang L, Yang TZ, Ji ZM. Applicability of the positive and negative affect scale in Chinese. Chin J Ment Health. 2003;17:54–6.

Imbir KK. (2015). Affective norms for 1,586 Polish words (ANPW): Duality-of-mind approach. *Behav Res Methods*. 2015 Sep; 47(3):860-70.

Imbir KK (2016). Affective Norms for 4900 Polish Words Reload (ANPW_R): Assessments for Valence, Arousal, Dominance, Origin, Significance, Concreteness, Imageability and, Age of Acquisition. Front. Psychol. 7:1081. doi: 10.3389/fpsyg.2016.01081

Izura, C. & Ellis, A. (2002). Age of acquisition effects in word recognition and production in first and second languages. *Psicológica*, 23, 245-281. www.uv.es/revispsi/articulos2.02/4.IZURA%26ELLIS.pdf

Jiang, K., Calix, R., Gupta, M. (2016). Construction of a Personal Experience Tweet Corpus for Health Surveillance. *Proc. 15th Workshop on Biomedical Natural Language Processing*, 128–135, Berlin, Germany, August 12, 2016. Association for Computational Linguistics. http://www.aclweb.org/anthology/W/W16/W16-2917.pdf

Judge, T., & Larsen, R. (2001). Dispositional Affect and Job Satisfaction: A Review and Theoretical Extension. *Organizational Behavior and Human Decision Processes*, 86(1), 67–98. doi:10.1006/obhd.2001.2973.

Karmiloff-Smith, A. (2009). Nativism versus neuroconstructivism: rethinking the study of developmental disorders. *Developmental Psychology*, 45(1), 56-63.

Kuperman, V. (2015). Virtual experiments in megastudies: a case study of language and emotion. Q. *J. Exp. Psychol.* 68, 1693–1710. doi: 10.1080/17470218.2014.989865

Kuppens, P., Tuerlinckx, F., Russell, J. A., & Barrett, L. F. (2013). The relation between valence and arousal in subjective experience. *Psychol Bull*, 139 (4), 917-940. doi:10.1037/a0030811

Lahti, Lauri (2016). Supporting diagnostics and decision making in healthcare by modular methods of computational linguistics. *Proc. E-Learn 2016 - World Conference on E-Learning*, 14–16 November 2016, Washington, D.C., USA (eds. Ho, C., & Lin, G.), 1513-1519. Association for the Advancement of Computing in Education (AACE), Chesapeake, VA, USA. ISBN 978-1-939797-25-4. https://www.learntechlib.org/p/174196

Lahti, Lauri, Tenhunen, H., Heinonen, S., Helkavaara, M., Pöyhönen-Alho, M., & Torkki, P. (2017). Data for: Development of computational models for emotional diary text analysis to support maternal care. Mendeley Data, v1. doi:10.17632/mm6h6mkkwy.1

Lang, P.J. (1980). Behavioral treatment and bio-behavioral assessment: computer applications. In J.B. Sidowski, J.H. Johnson, & T.A. Williams (eds.), Technology in mental healthcare delivery systems (pp. 119-137). Norwood, NJ: Ablex.

Lang, P.J., Bradley, M.M., & Cuthbert, B.N. (2008). International affective picture system (IAPS): Affective ratings of pictures and instruction manual. Technical Report A-8. University of Florida, Gainesville, FL.

Lee, E., Denison, F., Hor, K., & Reynolds, R. (2016). Web-based interventions for prevention and treatment of perinatal mood disorders: a systematic review. BMC Pregnancy and Childbirth (2016) 16:38. https://bmcpregnancychildbirth.biomedcentral.com/articles/10.1186/s12884-016-0831-1

Lever Taylor B, Cavanagh K, Strauss C (2016) The Effectiveness of Mindfulness-Based Interventions in the Perinatal Period: A Systematic Review and Meta-Analysis. PLoS ONE 11(5): e0155720. https://doi.org/10.1371/journal.pone.0155720

Lloyd, S. P. (1957). Least square quantization in PCM. Bell Telephone Laboratories Paper. Published in journal much later: Lloyd., S. P. (1982). "Least squares quantization in PCM" (PDF). *IEEE Transactions on Information Theory*. 28 (2): 129–137. doi:10.1109/TIT.1982.1056489.

McNair, D., Lorr, M., Droppleman, L. (1971). Edits Manual for the Profile of Mood States. San Diego: Educational and Industrial Testing Service; 1971.

Mehrabian, A. (1972). Nonverbal communication. In J. K. Cole (Eds.), Nebreska symposium on motivation, 1971. Vol.19, (pp 107-161), Lincoln: University of Nebreska Press.

Mehrabian, A., & Russell, J.A. (1974). An approach to environmental psychology. Cambridge, MA, USA; London, UK: MIT Press.

Monnier C1, Syssau A2. (2016). Affective norms for 720 French words rated by children and adolescents (FANchild). Behav Res Methods. 2016 Dec 7. [Epub ahead of print]

Montefinese M, Ambrosini E, Fairfield B, Mammarella N (2014). The adaptation of the Affective Norms for English Words (ANEW) for Italian. *Behav Res Methods*. 2014 Sep; 46(3):887-903.

Moors A, De Houwer J, Hermans D, Wanmaker S, van Schie K, Van Harmelen AL, De Schryver M, De Winne J, Brysbaert M (2013). Norms of valence, arousal, dominance, and age of acquisition for 4,300 Dutch words. *Behav Res Methods*. 2013 Mar; 45(1):169-77.

Nie, C., Dai, Q., Zhao, R., Dong, Y., Chen, Y. & Ren, H. (2017). The impact of resilience on psychological outcomes in women with threatened premature labor and spouses: a cross-sectional study in Southwest China. Health and Quality of Life Outcomes, 2017, 15:26. https://doi.org/10.1186/s12955-017-0603-2 https://hqlo.biomedcentral.com/articles/10.1186/s12955-017-0603-2

Osgood, C. (1952). The nature and measurement of meaning. *Psychological Bulletin*, 49, 172-237.

Parker, J., Wei, Y., Yates, A., Frieder, O., & Goharian. N. (2013). A Framework for Detecting Public Health Trends with Twitter. *Proc. IEEE/ACM International Conference on Advances in Social Networks Analysis and Mining*, 556-563. http://ir.cs.georgetown.edu/publications/downloads/parker-twitterhealth.pdf

Parker, R. M., Dmitrieva, E., Frolov, S., & Gazmararian, J. A. (2012). Text4baby in the United States and Russia: An opportunity for understanding how mHealth affects maternal and child health. *Journal of Health Communication*, 17(Suppl. 1), 30–36.

Parole corpus (2017). Parole corpus of Finnish written language. Institute for the languages of Finland. http://kaino.kotus.fi/sanat/taajuuslista/parole.php

Paul, M., & Dredze, M. (2012). A model for mining public health topics from twitter. *HEALTH*, 11:16–6. http://www.cs.jhu.edu/~mpaul/files/2011.tech.twitter_health.pdf

Posner, J., Russell, J. A., & Peterson, B. S. (2005). The circumplex model of affect: An integrative approach to affective neuroscience, cognitive development, and psychopathology. Development and Psychopathology, 17(3), 715–734. http://doi.org/10.1017/S0954579405050340

Pratto, F., and John, O. P. (1991). Automatic vigilance: the attention-grabbing power of negative social information. *J. Pers. Soc. Psychol.* 61, 380–391. doi: 10.1037/0022-3514.61.3.380

R Project (2017). The R Project for Statistical Computing. https://www.r-project.org/

R script scale method (2017). Scaling and Centering of Matrix-like Objects. Online documentation referring to the source: Becker, R. A., Chambers, J. M. and Wilks, A. R. (1988) The New S Language. Wadsworth & Brooks/Cole. http://stat.ethz.ch/R-manual/R-devel/library/base/html/scale.html

Redondo J1, Fraga I, Padrón I, Comesaña M. (2007). The Spanish adaptation of ANEW (affective norms for English words). Behav Res Methods. 2007 Aug;39(3):600-5.

Rennung M, & Göritz AS (2015) Facing Sorrow as a Group Unites. Facing Sorrow in a Group Divides. *PLoS ONE* 10(9): e0136750. doi:10.1371/journal.pone.0136750

Rosebrock, L., Hoxha, D., & Gollan, J. (2015). Affective reactivity differences in pregnant and postpartum women. Psychiatry Res. 2015 June 30; 227(0): 179–184. doi:10.1016/j.psychres.2015.04.002. http://pubmedcentralcanada.ca/pmcc/articles/PMC4430352/

Russell, J.A. (1980). A circumplex model of affect. *Journal of Personality and Social Psychology*, 39, 1161-1178.

Schlosberg, H. (1954). Three dimensions of emotion. *Psychological Review*, 61, 81-88.

Schmidtke DS1, Schröder T, Jacobs AM, Conrad M. (2014). ANGST: affective norms for German sentiment terms, derived from the affective norms for English words. Behav Res Methods. 2014 Dec;46(4):1108-18. doi: 10.3758/s13428-013-0426-y.

Scott, G. G., O'Donnell, P. J., and Sereno, S. C. (2014). Emotion words and categories: evidence from lexical decision. *Cogn. Process.* 15, 209–215. doi: 10.1007/s10339-013-0589-6

Sereno S, Scott G, Yao B, Thaden E, O'Donnell P (2015). Emotion word processing: does mood make a difference? *Front. Psychol.*, 24 August 2015 | https://doi.org/10.3389/fpsyg.2015.01191

Seth, A. K. (2013). Interoceptive inference, emotion, and the embodied self. *Trends Cogn Sci*, 17(11), 565-573. doi:10.1016/j.tics.2013.09.007

Sianipar A, van Groenestijn P and Dijkstra T (2016). Affective Meaning, Concreteness, and Subjective Frequency Norms for Indonesian Words. Front. Psychol. 7:1907. doi: 10.3389/fpsyg.2016.01907

Soares AP1, Comesaña M, Pinheiro AP, Simões A, Frade CS. (2012). The adaptation of the Affective Norms for English Words (ANEW) for European Portuguese. Behav Res Methods. 2012 Mar;44(1):256-69. doi: 10.3758/s13428-011-0131-7.

Swanson, L. W. (2012). Brain architecture: understanding the basic plan. Oxford University Press.

Söderholm C, Häyry E, Laine M, Karrasch M (2013) Valence and Arousal Ratings for 420 Finnish Nouns by Age and Gender. *PLoS ONE* 8(8): e72859. doi:10.1371/journal.pone.0072859 (http://journals.plos.org/plosone/article?id=10.1371/journal.pone.0072859)

Taylor, S. E. (1991). Asymmetrical effects of positive and negative events: the mobilization-minimization hypothesis. *Psychol. Bull.* 110, 67–85. doi: 10.1037/0033-2909.110.1.67

Tibshirani, R., Walther, G., & Hastie, T. (2001). Estimating the number of clusters in a data set via the gap statistic. *J. R. Statist. Soc. B* (2001), 63, Part 2, pp. 411-423.

Toivonen R, Kivelä M, Saramäki J, Viinikainen M, Vanhatalo M, Sams M (2012) Networks of Emotion Concepts. PLoS ONE 7(1): e28883. doi:10.1371/journal.pone.0028883. http://journals.plos.org/plosone/article?id=10.1371/journal.pone.0028883

Tuovila, S. (2005). Such emotions. The semantics of emotion words in the Finnish language. (Published in Finnish: Kun on tunteet. Suomen kielen tunnesanojen semantiikkaa.) Faculty of Humanities, University of Oulu, Finland. http://jultika.oulu.fi/files/isbn9514278070.pdf

Unkelbach, C., Fiedler, K., Bayer, M., Stegmüller, M., and Danner, D. (2008). Why positive information is processed faster: the density hypothesis. *J. Pers. Soc. Psychol.* 95, 36–49. doi: 10.1037/0022-3514.95.1.36

Van den Bergh, B.R.H, Marcoen, A., 2004. High antenatal maternal anxiety is related to ADHD symptoms, externalizing problems, and anxiety in 8 and 9 year olds. Child. Dev. 13, 1085-1097.

Wadhwa PD, Entringer S, Buss C, Lu MC (2011). The contribution of maternal stress to preterm birth: issues and considerations. Clin Perinatol. 2011 Sep; 38(3):351-84.

Warriner AB, Kuperman V, Brysbaert M (2013). Norms of valence, arousal, and dominance for 13,915 English lemmas. *Behav Res Methods*. 2013 Dec; 45(4):1191-207. (http://link.springer.com/article/10.3758/s13428-012-0314-x)

Watson D, Clark LA, Tellegen A. Development and validation of brief measures of positive and negative affect: the PANAS scales. J Pers Soc Psychol. 1988;54(6):1063–70.

Wundt, W. (1896). Outlines of psychology. Englemann, Leipzig, Germany.

Xu, F., & Kushnir, T. (2013). Infants are rational constructivist learners. *Current Directions in Psychological Science*, 22(1), 28-32.

# TABLES AND FIGURES

**Table 1. Some background information about the group of 35 persons of the experimental evaluation.**

| Group | Average age (standard deviation) | Completed Bachelor degree |
|---|---|---|
| All (n=35) | 30.83 (10.92) | 57 % |
| Women (n=21) | 31.95 (12.75) | 67 % |
| Men (n=14) | 29.14 (7.52) | 43 % |
| Women without children (n=13) | 25.46 (6.25) | 54 % |
| Women with children (n=8) | 42.50 (13.86) | 88 % |
| Men without children (n=13) | 28.77 (7.69) | 38 % |
| Men with children (n=1) | 34.00 (not defined) | 100 % |

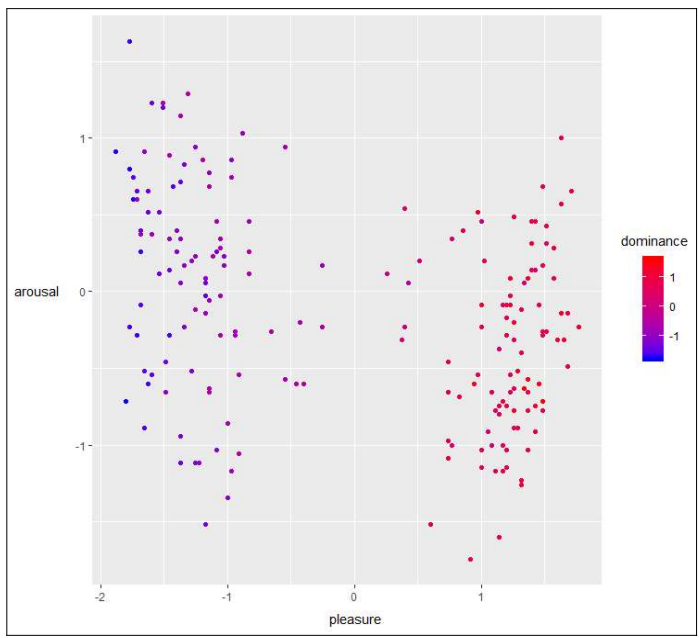

**Figure 1.** The distribution of average vectors of emotional adjectives for all 195 emotional adjectives (n=35). The dimension pleasure shown on x-axis, the dimension arousal on y-axis and the dimension dominance as a color ranging from blue to red.

**Table 2.** Average values and standard deviations for the affective rating answers for all persons and subgroups for 195 emotional adjectives (values shown computed as measures of average and standard deviation based on values given by persons belonging to the subgroup in question).

| Group | Pleasure | | Arousal | | Dominance | |
|---|---|---|---|---|---|---|
| | *average* | *standard deviation* | *average* | *standard deviation* | *average* | *standard deviation* |
| All (n=35) | -0.04 | 1.46 | -0.14 | 1.32 | -0.16 | 1.31 |
| Women (n=21) | -0.06 | 1.52 | -0.17 | 1.32 | -0.19 | 1.36 |
| Men (n=14) | 0 | 1.37 | -0.08 | 1.33 | -0.13 | 1.22 |
| Women without children (n=13) | -0.06 | 1.55 | -0.12 | 1.35 | -0.18 | 1.35 |
| Women with children (n=8) | -0.06 | 1.46 | -0.27 | 1.26 | -0.21 | 1.38 |
| Men without children (n=13) | -0.01 | 1.36 | -0.11 | 1.35 | -0.13 | 1.22 |
| Men with children (n=1) | 0.1 | 1.43 | 0.3 | 1.01 | -0.15 | 1.13 |

**Table 3.** Some distinctive average vectors of emotional adjectives towards the eight corners of the emotional vector space with affective rating values more or less than zero, in six pair-wise comparison combinations between three gender-parental subgroups Women with children (wwc; n=8), Women without children (wwoc; n=13) and Men without children (mwoc; n=13). Emotional adjectives are listed in the decreasing order of distance (shown in parenthesis).

| Eight corners of the emotional vector space with affective rating values more or less than zero | | | | | |
|---|---|---|---|---|---|
| | | dominance <0 | | dominance >0 | |
| | group | arousal <0 | arousal >0 | arousal <0 | arousal >0 |
| pleasure <0 | wwc & ¬wwoc: | uncomfortable (1.18) shared sense of shame (1.1) awkward (1.06) | disappointed (0.87) offended (0.78) inadequate (0.65) | No value | No value |
| | ¬wwc & wwoc: | small (1.01) disappointed (0.87) negative (0.74) | aggressive (1.18) uncomfortable (1.18) shared sense of shame (1.1) | No value | No value |
| | wwoc & ¬mwoc: | burned out (1.54) disappointed (1.27) stuck (0.91) | lost (1.79) aggressive (1.57) skeptical (1.21) | No value | No value |
| | ¬wwoc & mwoc: | lost (1.79) insecure (1.06) irresolute (1.05) | burned out (1.54) disappointed (1.27) stuck (0.91) | No value | No value |
| | wwc & ¬mwoc: | gloating (1.15) burned out (1.01) helpless (0.95) | terrible (1.47) dissatisfied (0.83) lost (0.81) | No value | No value |
| | ¬wwc & mwoc: | small (1.19) dissatisfied (0.83) lost (0.81) | gloating (1.15) burned out (1.01) judgmental (1) | No value | No value |
| pleasure >0 | wwc & ¬wwoc: | astonished (0.88) expecting (0.6) | touched (0.41) | ecstatic (1.02) small (1.01) alert (0.94) | productive (0.73) blissful (0.64) |
| | ¬wwc & wwoc: | kind (0.52) | naked (1.53) astonished (0.88) blissful (0.64) | productive (0.73) happy (0.59) | euphoric (1.73) ecstatic (1.02) alert (0.94) |
| | wwoc & ¬mwoc: | kind (0.85) | naked (1.45) astonished (0.49) | special (1.53) intrested (0.95) lovely (0.82) | capable (0.86) cheerful (0.85) ecstatic (0.84) |
| | ¬wwoc & mwoc: | No value | ecstatic (0.84) touched (0.46) | expecting (0.88) capable (0.86) kind (0.85) | naked (1.45) interested (0.95) lovely (0.82) |
| | wwc & ¬mwoc: | astonished (1.08) expecting (0.5) | No value | ecstatic (1.25) special (1.22) small (1.19) | productive (1.09) blissful (0.86) efficient (0.63) |
| | ¬wwc & mwoc: | No value | ecstatic (1.25) blissful (0.86) | productive (1.09) happy (0.6) expecting (0.5) | naked (2.04) euphoric (1.66) energetic (1.24) |

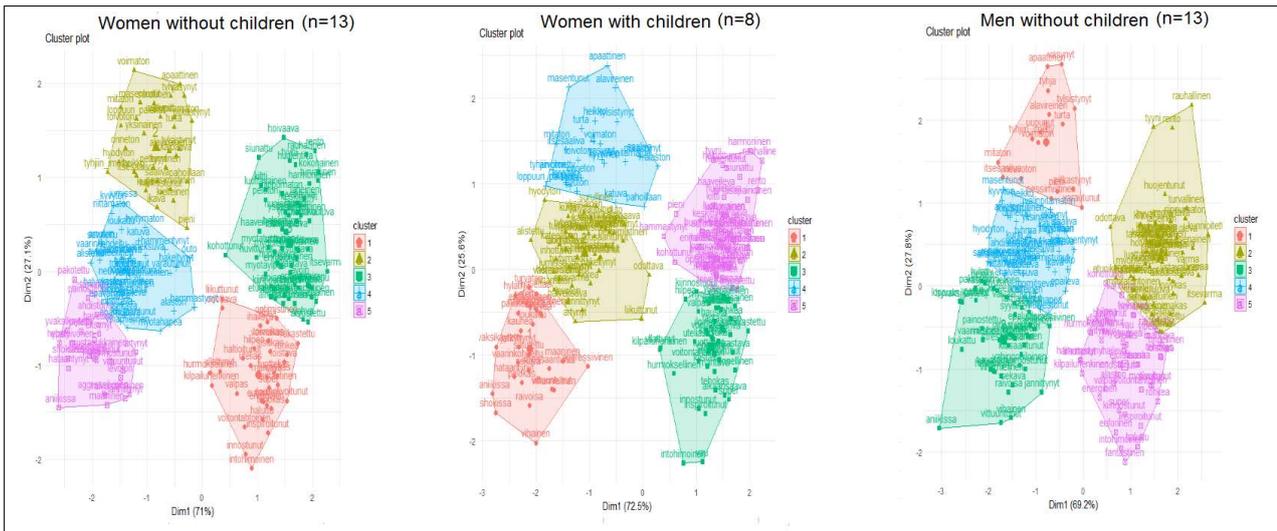

**Figure 2.** Five clusters of affective rating data generated with k-means clustering method for gender-parental subgroups. The visualization is a projection from a three-dimensional space to a two-dimensional space.

**Table 4.** The five average vectors of clusters for each gender-parental subgroup in respect to the five corresponding most matching clusters between subgroups.

| Women without children (n=13) | | | | | Women with children (n=8) | | | | | Men without children (n=13) | | | | |
|---|---|---|---|---|---|---|---|---|---|---|---|---|---|---|
| cluster name | pleasure | arousal | dominance | distance from the origo | cluster name | pleasure | arousal | dominance | distance from the origo | cluster name | pleasure | arousal | dominance | distance from the origo |
| A_Women_without_children | 1.248 | 0.238 | 0.738 | 1.469 | A_Women_with_children | 1.321 | 0.069 | 0.867 | 1.583 | A_Men_without_children | 1.232 | 0.153 | 0.627 | 1.391 |
| B_Women_without_children | -1.218 | -0.685 | -1.075 | 1.763 | B_Women_with_children | -1.231 | -0.702 | -1.174 | 1.840 | B_Men_without_children | -1.163 | -0.854 | -1.104 | 1.816 |
| C_Women_without_children | 1.202 | -0.709 | 0.791 | 1.604 | C_Women_with_children | 1.115 | -0.744 | 0.688 | 1.507 | C_Men_without_children | 1.177 | -0.754 | 0.842 | 1.631 |
| D_Women_without_children | -1.133 | 0.209 | -0.978 | 1.511 | D_Women_with_children | -1.042 | 0.101 | -0.891 | 1.374 | D_Men_without_children | -1.227 | -0.100 | -1.046 | 1.615 |
| E_Women_without_children | -1.399 | 0.866 | -1.148 | 2.006 | E_Women_with_children | -1.498 | 0.838 | -1.192 | 2.090 | E_Men_without_children | -1.337 | 0.676 | -1.080 | 1.847 |

**Table 5.** Three closest average vectors of emotional adjectives for each average vector of a cluster in respect to five clusters for each gender-parental subgroup. Notation: emot. adj. = emotional adjective, distance from the avg vector of the cluster = distance of the average vector of the emotional adjective from the average vector of the cluster, distance from the origo = distance of the average vector of the emotional adjective from the origo, cosine btw the avg vectors of the cluster and the emot. adj. = cosine similarity measure between the average vector of the cluster and the average vector of the emotional adjective.

### Women without children (n=13)

**A_Women_without_children**

| emot. adj. | distance from the avg vector of the cluster | distance from the origo | cosine btw the avg vectors of the cluster and the emot. adj. |
|---|---|---|---|
| curious | 0.15 | 1.46 | 0.99 |
| admiring | 0.24 | 1.41 | 0.99 |
| gritty | 0.31 | 1.72 | 0.99 |

**B_Women_without_children**

| emot. adj. | distance from the avg vector of the cluster | distance from the origo | cosine btw the avg vectors of the cluster and the emot. adj. |
|---|---|---|---|
| disappointed | 0.12 | 1.88 | 1 |
| irritated | 0.27 | 1.59 | 0.99 |
| miserable_I | 0.29 | 1.86 | 0.99 |

**C_Women_without_children**

| emot. adj. | distance from the avg vector of the cluster | distance from the origo | cosine btw the avg vectors of the cluster and the emot. adj. |
|---|---|---|---|
| committed | 0.1 | 1.63 | 1 |
| considerate | 0.17 | 1.6 | 0.99 |
| positive | 0.18 | 1.54 | 0.99 |

**D_Women_without_children**

| emot. adj. | distance from the avg vector of the cluster | distance from the origo | cosine btw the avg vectors of the cluster and the emot. adj. |
|---|---|---|---|
| bothered | 0.12 | 1.51 | 1 |
| disapproving | 0.14 | 1.48 | 1 |
| gloating | 0.17 | 1.46 | 0.99 |

**E_Women_without_children**

| emot. adj. | distance from the avg vector of the cluster | distance from the origo | cosine btw the avg vectors of the cluster and the emot. adj. |
|---|---|---|---|
| uncomfortable | 0.18 | 1.87 | 1 |
| tense_II | 0.3 | 1.88 | 0.99 |
| ashamed | 0.34 | 2.33 | 1 |

### Women with children (n=8)

**A_Women_with_children**

| emot. adj. | distance from the avg vector of the cluster | distance from the origo | cosine btw the avg vectors of the cluster and the emot. adj. |
|---|---|---|---|
| powerful | 0.2 | 1.64 | 0.99 |
| wonderful | 0.23 | 1.81 | 1 |
| proud | 0.24 | 1.71 | 0.99 |

**B_Women_with_children**

| emot. adj. | distance from the avg vector of the cluster | distance from the origo | cosine btw the avg vectors of the cluster and the emot. adj. |
|---|---|---|---|
| lonely | 0.34 | 1.81 | 0.98 |
| bothered | 0.35 | 1.64 | 0.99 |
| tired | 0.35 | 1.91 | 0.98 |

**C_Women_with_children**

| emot. adj. | distance from the avg vector of the cluster | distance from the origo | cosine btw the avg vectors of the cluster and the emot. adj. |
|---|---|---|---|
| compassionate | 0.06 | 1.55 | 1 |
| understanding | 0.14 | 1.49 | 1 |
| positive | 0.19 | 1.59 | 0.99 |

**D_Women_with_children**

| emot. adj. | distance from the avg vector of the cluster | distance from the origo | cosine btw the avg vectors of the cluster and the emot. adj. |
|---|---|---|---|
| tense_II | 0.15 | 1.26 | 1 |
| lost | 0.24 | 1.51 | 0.99 |
| disappointed | 0.24 | 1.51 | 0.99 |

**E_Women_with_children**

| emot. adj. | distance from the avg vector of the cluster | distance from the origo | cosine btw the avg vectors of the cluster and the emot. adj. |
|---|---|---|---|
| hurt | 0.28 | 2.35 | 1 |
| guilty | 0.31 | 2.22 | 0.99 |
| disoriented | 0.34 | 2.33 | 0.99 |

### Men without children (n=13)

**A_Men_without_children**

| emot. adj. | distance from the avg vector of the cluster | distance from the origo | cosine btw the avg vectors of the cluster and the emot. adj. |
|---|---|---|---|
| funny | 0.01 | 1.39 | 1 |
| proud | 0.24 | 1.41 | 0.99 |
| motivated | 0.33 | 1.71 | 1 |

**B_Men_without_children**

| emot. adj. | distance from the avg vector of the cluster | distance from the origo | cosine btw the avg vectors of the cluster and the emot. adj. |
|---|---|---|---|
| drained | 0.09 | 1.89 | 1 |
| exhausted | 0.14 | 1.84 | 1 |
| melancholic | 0.26 | 1.82 | 0.99 |

**C_Men_without_children**

| emot. adj. | distance from the avg vector of the cluster | distance from the origo | cosine btw the avg vectors of the cluster and the emot. adj. |
|---|---|---|---|
| harmonious | 0.1 | 1.55 | 1 |
| whole | 0.12 | 1.53 | 1 |
| present | 0.13 | 1.7 | 1 |

**D_Men_without_children**

| emot. adj. | distance from the avg vector of the cluster | distance from the origo | cosine btw the avg vectors of the cluster and the emot. adj. |
|---|---|---|---|
| miserable_II | 0.18 | 1.65 | 0.99 |
| ashamed | 0.2 | 1.7 | 0.99 |
| stupid | 0.22 | 1.44 | 0.99 |

**E_Men_without_children**

| emot. adj. | distance from the avg vector of the cluster | distance from the origo | cosine btw the avg vectors of the cluster and the emot. adj. |
|---|---|---|---|
| mistreated | 0.13 | 1.94 | 1 |
| jealous | 0.25 | 1.6 | 1 |
| shocked | 0.31 | 1.72 | 0.99 |

**Table 6.** The 16 average vectors of a pregnancy-related noun for each gender-parental subgroup and the nearest average vectors of clusters in an ascending order in respect to the distance between the average vector of the pregnancy-related noun and the average vector of the cluster for the subgroup in question.

| pregnancy-related noun | Women without children (n=13) | | | | | Women with children (n=8) | | | | | Men without children (n=13) | | | | |
|---|---|---|---|---|---|---|---|---|---|---|---|---|---|---|---|
| | pleasure | arousal | dominance | distance from origo | the nearest average vectors of clusters in an ascending order of distance | pleasure | arousal | dominance | distance from origo | the nearest average vectors of clusters in an ascending order of distance | pleasure | arousal | dominance | distance from origo | the nearest average vectors of clusters in an ascending order of distance |
| intimate relationship | 1.31 | -0.46 | 0.54 | 1.46 | C, A, D, B, E | 1.13 | -0.13 | 1.13 | 1.14 | A, C, D, B, E | 0.85 | -0.23 | 0.46 | 0.91 | A, C, D, B, E |
| motherhood | 0.62 | 0.23 | -0.23 | 0.7 | A, C, D, B, E | 1.75 | 0 | 1.13 | 1.75 | A, C, D, B, E | 0.62 | -0.31 | 0 | 0.75 | A, C, D, B, E |
| fatherhood | 0.69 | -0.54 | 0.31 | 1.03 | C, A, D, B, E | 1.38 | -0.25 | 1 | 1.42 | A, C, D, B, E | 1 | -0.77 | 0.23 | 1.48 | C, A, B, D, E |
| infant | 0.85 | -0.15 | -0.23 | 0.87 | A, C, D, B, E | 1.63 | -0.25 | 0.88 | 1.66 | A, C, D, B, E | 0.62 | -0.69 | -0.23 | 1.16 | C, A, B, D, E |
| fetus | -0.31 | 0.31 | -0.54 | 0.53 | D, E, B, A, C | 1 | 0 | 0 | 1 | A, C, D, B, E | 0.23 | -0.54 | -0.69 | 0.8 | B, D, A, C, E |
| pregnancy | 0.39 | 0.54 | -0.31 | 0.85 | A, D, C, E, B | 1.13 | 0.25 | 0.25 | 1.18 | A, C, D, B, E | 0.46 | -0.39 | -0.39 | 0.71 | A, C, D, B, E |
| giving birth | -0.39 | 0.77 | -1.15 | 1.15 | D, E, B, A, C | 0.25 | 1.13 | 0 | 1.61 | A, D, E, C, B | 0.15 | 0.31 | -0.85 | 0.46 | D, E, B, A, C |
| breastfeeding | 0.69 | -0.46 | 0.08 | 0.95 | C, A, D, B, E | 1.13 | -0.75 | 0.25 | 1.55 | C, A, D, B, E | 0.54 | -0.62 | 0.08 | 1.02 | C, A, B, D, E |
| baby colic | -1.46 | 0.92 | -1.39 | 1.96 | E, D, B, A, C | -1.13 | 0.25 | -1.38 | 1.18 | D, E, B, C, A | -1.08 | 0.54 | -0.85 | 1.32 | E, D, B, A, C |
| miscarriage | -1.92 | 0.85 | -1.92 | 2.27 | E, D, B, A, C | -2 | 1.13 | -2 | 2.56 | E, D, B, A, C | -1.62 | 0.92 | -1.69 | 2.08 | E, D, B, A, C |
| abortion | -1.15 | 0.39 | -0.54 | 1.28 | D, E, B, A, C | -1.25 | 0.13 | -1 | 1.26 | D, E, B, C, A | -0.77 | 0.31 | -0.54 | 0.88 | D, E, B, A, C |
| preemie | -0.92 | 0.54 | -1.31 | 1.2 | D, E, B, A, C | -0.25 | 0.38 | -0.88 | 0.59 | D, E, B, C, A | -0.69 | -0.08 | -1.08 | 0.7 | D, B, E, A, C |
| childlessness | -1.31 | -0.62 | -1.39 | 1.57 | B, D, E, A, C | -1.5 | -0.38 | -1.63 | 1.59 | B, D, E, C, A | -0.85 | -0.46 | -0.46 | 1.07 | D, B, E, C, A |
| sexuality | 1.46 | 0.39 | 0.54 | 1.56 | A, C, D, B, E | 1.13 | 0.25 | 0.63 | 1.18 | A, C, D, B, E | 1.62 | 0.39 | 0.39 | 1.71 | A, C, D, E, B |
| sole custody of child | -0.77 | -0.08 | -0.54 | 0.78 | B, E, A, C | 0 | -0.38 | 0 | 0.53 | C, D, A, B, E | -0.62 | -0.08 | -0.69 | 0.63 | D, B, E, A, C |
| artificial fertilization | -0.08 | -0.08 | -0.31 | 0.13 | D, B, A, C, E | 0.63 | -1 | -0.13 | 1.55 | C, A, D, B, E | 0.15 | -0.31 | -0.62 | 0.46 | D, B, A, C, E |

**Table 7.** Emerging patterns of affectivity of subgroups, words or clusters in respect to the proposed models when aiming to support maternal care. Notation: wwoc = women without children, wwc = women with children and mwoc = men without children.

| Model | Emerging patterns |
|---|---|
| General transformation model | Subgroups having a general transformation vector with the greatest distance from the origo:<br><br>For gender subgroups: $T_{women \rightarrow men}$ (0.124). For gender-parental subgroups: $T_{wwc \rightarrow mwoc}$ (0.186). For rating-daytime subgroups: $T_{middle \rightarrow late}$ (0.302). For rating-duration subgroups: $T_{short \rightarrow long}$ (0.262). |
| Word-specific transformation model (emotional adjectives) | Words having the greatest distances between gender-parental subgroups:<br><br>wwoc vs. wwc: euphoric (1.735); naked (1.533); terrible (1.191); aggressive (1.185); uncomfortable (1.180).<br>wwoc vs. mwoc: lost (1.791); aggressive (1.573); burned out (1.542); special (1.527); naked (1.451).<br>wwc vs. mwoc: naked (2.037); euphoric (1.656); wow! (1.511); terrible (1.469); appreciative (1.399). |
| Extreme affective shift model | Words of the current gender-parental subgroup identified to have the greatest shift in comparison with one of the other two gender-parental subgroups:<br><br>wwc & ¬wwoc: uncomfortable (1.18); shared sense of shame (1.1); awkward (1.06); ecstatic (1.02); small (1.01)<br>¬wwc & wwoc: euphoric (1.73); naked (1.53); aggressive (1.18); uncomfortable (1.18); shared sense of shame (1.1)<br>wwoc & ¬mwoc: lost (1.79); aggressive (1.57); burned out (1.54); special (1.53); naked (1.45)<br>¬wwoc & mwoc: lost (1.79); burned out (1.54); naked (1.45); disappointed (1.27); insecure (1.06)<br>wwc & ¬mwoc: terrible (1.47); ecstatic (1.25); special (1.22); small (1.19); gloating (1.15)<br>¬wwc & mwoc: naked (2.04); euphoric (1.66); ecstatic (1.25); energetic (1.24); small (1.19) |
| Cluster-based transformation model | The greatest distances between the matching clusters of the gender-parental subgroups:<br><br>wwoc vs. wwc: Cluster A (0.225); Cluster D (0.166); Cluster C (0.139); Cluster E (0.112); Cluster B (0.101).<br>wwoc vs. mwoc: Cluster D (0.330); Cluster E (0.211); Cluster B (0.180); Cluster A (0.141); Cluster C (0.072).<br>wwc vs. mwoc: Cluster D (0.314); Cluster A (0.270); Cluster E (0.254); Cluster B (0.181); Cluster C (0.166). |
| Scale-cluster model | Words shared by the lists of scale-cluster models of two gender-parental subgroups, supplied with the substraction of distances from the origo:<br><br>wwoc & wwc: Cluster A: gritty (-0.197); amazing (-0.185). Cluster B: miserable_II (0.43); depressed (-0.484). Cluster C: compassionate (-0.157); positive (-0.046); present (0.154); appreciative (-0.237). Cluster D: envious (-0.039). Cluster E: disoriented (-0.602).<br>wwoc & mwoc: Cluster A: curious (0.458). Cluster B: No value. Cluster C: committed (0.052); present (0.1); respectful (0.634); appreciative (1.153). Cluster D: disapproving (0.323); stupid (0.464); worthless (0.583); hurt (0.449). Cluster E: tense_II (0.829); jealous (0.672); stressed (0.88).<br>wwc & mwoc: Cluster A: wonderful (-0.052). Cluster B: pessimistic (-0.335). Cluster C: understanding (-0.142); present (-0.054); open-minded (0.349). Cluster D: disgusting (0.842). Cluster E: annoyed (0.09); offended (0.051); mistreated (0.622). |
| Word-specific transformation model (pregnancy-related nouns) | Words having the greatest distances between gender-parental subgroups:<br><br>wwoc vs. wwc: motherhood (1.783); fetus (1.450); giving birth (1.364); infant (1.360); artificial fertilization (1.176).<br>wwoc vs. mwoc: childlessness (1.050); fetus (1.018); pregnancy (0.936); giving birth (0.770); baby colic (0.762).<br>wwc vs. mwoc: motherhood (1.628); infant (1.564); childlessness (1.341); giving birth (1.185); fetus (1.166). |
| Attraction-cluster model | Pregnancy-related nouns having different nearest clusters among matching clusters of gender-parental subgroups:<br><br>wwoc vs. wwc: intimate relationship (C vs. A); fatherhood (C vs. A); fetus (D vs. A); giving birth (D vs. A); baby colic (E vs. D); sole custody of child (D vs. C); artificial fertilization (D vs. C)<br>wwoc vs. mwoc: intimate relationship (C vs. A); infant (A vs. C); fetus (D vs. B); childlessness (B vs. D)<br>wwc vs. mwoc: fatherhood (A vs. C); infant (A vs. C); fetus (A vs. B); giving birth (A vs. D); baby colic (D vs. E); childlessness (B vs. D); sole custody of child (C vs. D); artificial fertilization (C vs. D) |

zzzzzzz

**Article title:** APPENDICES FOR RESEARCH ARTICLE "Development of computational models for emotional diary text analysis to support maternal care"

**Corresponding author:** Lauri Lahti (email: lauri.lahti@aalto.fi)

*Supplementary Material*

# Development of computational models for emotional diary text analysis to support maternal care

**Lauri Lahti**[*]**, Henni Tenhunen, Seppo Heinonen, Minna Helkavaara, Maritta Pöyhönen-Alho, Paulus Torkki**

Correspondence:

Lauri Lahti

lauri.lahti@aalto.fi

## Appendix A

**Table A1.** The set of 195 emotional adjectives in Finnish language used in the experimental evaluation listed here in the original order used in the rating task. Each Finnish concept is supplied with a unique coarse English translation. Additionally this table lists matching lemmas of Söderholm et al. (2013) and Eilola & Havelka (2010) and matching lemmas of those lemmas of Tuovila (2006) that have frequency of at least 2.

| 195 emotional adjectives | | Matching lemmas | |
|---|---|---|---|
| *Finnish concepts used in the experimental evaluation* | *Unique coarse English translations of corresponding Finnish concepts* | *Matching lemmas of Söderholm et al. (2013) (S_et_al) and Eilola & Havelka (2010) (E&H)* | *Matching lemmas of those lemmas of Tuovila (2006) that have frequency of at least 2* |
| aggressiivinen | aggressive | | aggressio |
| ahdistunut | anxious | E&H: ahdistus<br>S_et_al: ahdistus | ahdistus |
| aikaansaava | productive | | |
| ainutlaatuinen | unique | | |
| alaston | naked | | |
| alavireinen | melancholic | | |
| alistettu | oppressed | | alistuneisuus |
| apaattinen | apathetic | | apaattisuus |
| arvostava | appreciative | S_et_al: arvokkuus | |
| arvostettu | appreciated | S_et_al: arvokkuus | |
| arvoton | worthless | | |
| avoin | open-minded | | avoimuus |
| avuton | helpless | | avuttomuus |
| eksynyt | lost | | |
| eloisa | lively | E&H: elämä<br>S_et_al: elämä | |
| energinen | energetic | | energisyys |
| ennakoiva | anticipating | | |
| epäilevä | skeptical | | epäily(s) |
| epäluuloinen | distrustful | | epäluuloisuus |
| epämukava | uncomfortable | | |
| epätoivoinen | desperate | | epätoivo |
| epävarma | insecure | | epävarmuus |
| erityinen | special | | |
| etuoikeutettu | privileged | | |
| euforinen | euphoric | | |
| fantastinen | fantastic | | |
| haaveileva | dreaming | S_et_al: haaveilu | haaveilevuus |
| haltioitunut | ecstatic | | haltioituminen |
| haluttu | desired | E&H: halu | halu |
| halveksuva | contemptuous | | halveksunta |
| harmistunut | irritated | | harmi |
| harmoninen | harmonious | | |
| hauska | funny | | hauskuus |
| heikko | weak | | |
| helpottunut | relieved | | helpotus |
| hermostunut | nervous | | hermostuneisuus |
| hilpeä | cheerful | | hilpeys |
| hirveä | horrible | | |
| hoivaava | nurturing | S_et_al: hoitaja | |
| huojentunut | reassured | | |
| huolehtiva | caring | | huolehtivuus |
| huolestunut | concerned | | huoli |
| huomioiva | considerate | | |
| hurmoksellinen | blissful | S_et_al: hurmio | |
| huvittunut | amused | | huvittuneisuus |
| hylätty | abandoned | S_et_al: hylkäys | hylättyys |
| hyväksikäytetty | abused | | |
| hyväksyvä | accepting | E&H: hyväksyntä<br>S_et_al: hyvyys | |
| hyödytön | useless | | |
| häkeltynyt | overwhelmed | | |
| hämmentynyt | confused | | hämmennys |
| hämmästynyt | astonished | | hämmästys |
| hätääntynyt | alarmed | | hätä |
| hävettää | ashamed | S_et_al: häpeä | häpeä |
| ihaileva | admiring | S_et_al: ihailu | ihastus |
| ihana | lovely | S_et_al: ihanuus | ihana |
| ikävä | longing | | ikävä |
| ilahtunut | delighted | E&H: ilo<br>S_et_al: iloisuus | ilo |
| iloinen | joyful | E&H: ilo | ilo |

| | | S_et_al: iloisuus | |
|---|---|---|---|
| inhottava | disgusting | | inho |
| innostunut | excited | S_et_al: innokkuus | innostus |
| inspiroitunut | inspired | | |
| intohimoinen | passionate | S_et_al: intohimo | intohimo |
| itsenäinen | independent | | itsenäisyys |
| itsesäälivä | self-pitying | | |
| itsevarma | confident | | itsevarmuus |
| jumissa | stuck | | |
| jännittynyt | tense_I | | jännitys |
| järkyttynyt | upset | | |
| kaipaava | wishful | | kaipaus |
| kateellinen | envious | | kateus |
| katuva | remorseful | | katumus |
| kauhea | terrible | E&H: kauhu | |
| keskittynyt | focused | | |
| kielteinen | negative | | |
| kiinnostunut | interested | | kiinnostus |
| kiitollinen | grateful | S_et_al: kiitos | kiitollisuus |
| kilpailuhenkinen | competitive | | |
| kiltti | kind | S_et_al: kiltteys | |
| kireä | tense_II | | kireys |
| kiusaantunut | awkward | | kiusaantuneisuus |
| kohottunut | elevated | | |
| kokonainen | whole | | |
| kunnioitettu | respected | E&H: kunnioitus<br>S_et_al: kunnia | kunnioitus |
| kunnioittava | respectful | E&H: kunnioitus<br>S_et_al: kunnia | kunnioitus |
| kurja | miserable_I | E&H: kurjuus | |
| kyvykäs | capable | | |
| kyvytön | incapable | | |
| kyyninen | cynical | | kyynisyys |
| levoton | restless | | levottomuus |
| liikuttunut | touched | | |
| loistava | wonderful | S_et_al: loisto | |
| loppuun palanut | burned out | S_et_al: loppu | |
| loukattu | offended | E&H: loukkaus<br>S_et_al: loukkaus | loukkaantuneisuus |
| luottavainen | trusting | E&H: luottamus | luottamus |
| läheinen | close | S_et_al: läheisyys | läheisyys |
| lämmin | warm | E&H: lämpö<br>S_et_al: lämpö | lämpimyys&lämpö |
| läsnä oleva | present | | |
| maaninen | manic | | |
| mahtava | amazing | | |
| masentunut | depressed | E&H: masennus<br>S_et_al: masennus | masennus |
| mitätön | insignificant | | |
| motivoitunut | motivated | | |
| mustasukkainen | jealous | | mustasukkaisuus |
| myönteinen | positive | | |
| myötähäpeä | shared sense of shame | | |
| myötätuntoinen | compassionate | | myötätunto |
| myötäylpeä | shared sense of pride | | |
| neuvoton | irresolute | S_et_al: neuvonta | |
| nolostunut | embarrassed | | |
| odottava | expecting | | odotus |
| onnekas | lucky | E&H: onni | |
| onnellinen | happy | E&H: onni | onnellisuus |
| onneton | miserable_II | E&H: onnettomuus | onnettomuus |
| optimistinen | optimistic | | |
| osaava | competent | | |
| outo | weird | | |
| paheksuva | disapproving | | |
| pahoillaan | sorry | | |
| painostettu | pressured | S_et_al: painostus | |
| pakotettu | forced | | |
| paniikissa | panicked | | paniikki |
| pelokas | fearful | E&H: pelko<br>S_et_al: pelokkuus | pelko |
| peloton | fearless | E&H: pelko<br>S_et_al: pelokkuus | |
| pessimistinen | pessimistic | | |
| petetty | betrayed | E&H: petos<br>S_et_al: petos | petollisuus |
| pettynyt | disappointed | E&H: pettymys<br>S_et_al: pettymys | pettymys |
| pieni | small | | |
| pitkästynyt | dull | | pitkästyminen |
| raivoisa | furious | E&H: raivo<br>S_et_al: raivo | raivo |
| rakastava | loving | S_et_al: rakkaus | rakkaus |
| rakastettu | loved | S_et_al: rakkaus | |
| rauhallinen | calm | E&H: rauha<br>S_et_al: rauha | rauhallisuus |
| rento | relaxed | | rentous |
| riippumaton | autonomous | | |
| riittämätön | inadequate | | |
| rohkea | courageous | S_et_al: rohkeus | rohkeus |
| satutettu | hurt | | |
| sekava | disoriented | S_et_al: sekaannus | sekavuus |
| shokissa | shocked | | |
| sisukas | gritty | | |
| sitoutunut | committed | | |
| siunattu | blessed | | |
| stressaantunut | stressed | | stressi |
| super | superb | | |
| surullinen | sad | | suru |
| syyllinen | guilty | E&H: syyllisyys | syyllisyys |
| säälivä | pitying | | sääli |
| tasapainoinen | balanced | | tasapainoisuus |
| tehokas | efficient | S_et_al: tehokkuus | |
| toiveikas | hopeful | E&H: toivo<br>S_et_al: toive | toiveikkuus&toivo |
| toivoton | hopeless | E&H: toivo<br>S_et_al: toive | toivottomuus |
| tuomitseva | judgmental | S_et_al: tuomio | |
| tuottelias | prolific | E&H: tuotto<br>S_et_al: tuotos | |
| turhautunut | frustrated | | turhautuneisuus |
| turta | numb | | turtumus |
| turvallinen | safe | E&H: turvallisuus<br>S_et_al: turva | turvallisuus |
| turvaton | unsafe | E&H: turvallisuus<br>S_et_al: turva | |
| tyhjiin imetty | drained | S_et_al: tyhjyys | |
| tyhjä | empty | S_et_al: tyhjyys | tyhjyys |
| tyhmä | stupid | S_et_al: tyhmyys | |
| tylsistynyt | bored | | tylsyys |
| tyyni | serene | S_et_al: tyyneys | tyyneys |
| tyytymätön | dissatisfied | | tyytymättömyys |
| tyytyväinen | satisfied | | tyytyväisyys |
| täydellinen | perfect | E&H: täydellisyys | täydellisyys |
| utelias | curious | | uteliaisuus |
| uupunut | exhausted | | |
| vahingoniloinen | gloating | | |
| vahva | strong | | vahvuus |
| vaivaantunut | bothered | E&H: vaiva | vaivaantuneisuus |
| valpas | alert | | |
| vapaa | free | E&H: vapaus<br>S_et_al: vapaus | vapaus&vapautuneisuus |
| varautunut | reserved | | varautuneisuus |
| varma | assured | S_et_al: varmuus | varmuus |
| vastenmielinen | repulsive | | vastenmielisyys |
| vastuullinen | responsible | | |
| vau! | wow! | | |
| vihainen | angry | E&H: viha | viha |
| vihamielinen | hostile | | |
| vittuuntunut | pissed | | vitutus |
| voimakas | powerful | S_et_al: voima | voimakkuus |
| voimaton | powerless | S_et_al: voima | voimattomuus |
| voitontahtoinen | determined to win | | |
| välinpitämätön | indifferent | | välinpitämättömyys |
| välittävä | affectionate | | välittäminen |
| väsynyt | tired | | väsymys |
| väärinkohdeltu | mistreated | | |
| yhteenkuuluva | cohesive | | |
| yksinäinen | lonely | E&H: yksinäisyys | yksinäisyys |
| yllättynyt | surprised | | yllättyneisyys |
| ylpeä | proud | S_et_al: ylpeys | ylpeys |
| ymmärtävä | understanding | | ymmärtäväisyys |
| ystävällinen | friendly | E&H: ystävä<br>S_et_al: ystävyys | ystävällisyys&ystävyys&ystävä |
| ärtynyt | annoyed | | ärtymys&ärsyyntyminen |

**Table A2.** The set of 16 pregnancy-related nouns in Finnish language used in the experimental evaluation listed here in the original order used in the rating task. Each Finnish concept is supplied with a unique coarse English translation. Additionally this table lists matching lemmas of Söderholm et al. (2013) and Eilola & Havelka (2010).

| 16 pregnancy-related nouns | | Matching lemmas |
| --- | --- | --- |
| *Finnish concepts used in the experimental evaluation* | *Unique coarse English translations of corresponding Finnish concepts* | *Matching lemmas with Söderholm et al. (2013) (S_et_al) or Eilola & Havelka 2010 (E&H)* |
| parisuhde | intimate relationship | |
| äitiys | motherhood | E&H: isä |
| isyys | fatherhood | E&H: isä |
| vauva | infant (baby) | E&H: isä |
| sikiö | fetus | |
| raskaus | pregnancy | |
| synnytys | giving birth (delivery, childbirth) | S_et_al: syntymä |
| imetys | breastfeeding (lactation) | |
| koliikki | baby colic | |
| keskenmeno | miscarriage | |
| raskaudenkeskeytys | abortion (cessation of pregnancy) | S_et_al: abortti |
| keskonen | preemie (prematurely born child) | |
| lapsettomuus | childlessness | |
| seksuaalisuus | sexuality | S_et_al: seksi |
| yksinhuoltajuus | sole custody of child | |
| keinohedelmöitys | artificial fertilization | |

**Table A3.** In respect to 221 word lemmas having a frequency of at least two in the collection of Finnish emotion words (Tuovila 2005), those 110 lemmas of 221 lemmas that did not have a match with our 195 emotional adjective set, shown in a decreasing order of ranking position based on frequency (suffix -s indicates a shared ranking position).

*Finnish word lemmas (ranking position)*

tuska (14s), katkeruus (16s), riemu (16s), empatia (21s), hellyys (22s), kauhu (23s), himo (25s), kipu (25s), suuttumus (25s), ahneus (26s), apeus (27s), kylmyys (28s), pirteys (28s), usko (28s), anteeksianto (29s), arkuus (29s), kiukku (29s), lempeys (29s), levollisuus (29s), nauru (29s), rauhattomuus (29s), riehakkuus (29s), alakuloisuus (30s), hymy (30s), itku (30s), kaiho (30s), kärsivällisyys (30s), nälkä (30s), positiivisuus (30s), ujous (30s), haikeus (31s), hyvä (31s), kiiho (31s), kiihtyneisyys (31s), kiintymys (31s), kyllästyminen (31s), murhe (31s), murjotus (31s), nautinto (31s), onnistuneisuus (31s), rehellisyys (31s), riidanhaluisuus (31s), sympatia (31s), anteliaisuus (32s), epätietoisuus (32s), epäonni (32s), epäusko (32s), harmaus (32s), hartaus (32s), ihmetys (32s), ilkeys (32s), itsekkyys (32s), kauna (32s), kärsimys (32s), laiskuus (32s), onttous (32s), viileys (32s), ailahtelevuus (33s), aktiivisuus (33s), alemmuus (33s), aurinkoisuus (33s), avuliaisuus (33s), epämieltymys (33s), haluttomuus (33s), happamuus (33s), herkkyys (33s), huolettomuus (33s), ikävystyneisyys (33s), jano (33s), kaveruus (33s), kiire (33s), kostonhalu (33s), kovuus (33s), kuumuus (33s), kärsimättömyys (33s), lamaannus (33s), lannistuneisuus (33s), mielihyvä (33s), mukavuus (33s), negatiivisuus (33s), neutraalius (33s), pahantuulisuus (33s), paha olo (33s), passiivisuus (33s), paine (33s), penseys (33s), pitäminen (33s), sankarillisuus (33s), sieto (33s), sovinnollisuus (33s), surkeus (33s), suruttomuus (33s), synkkyys (33s), tsemppi (33s), tulistuminen (33s), tunteettomuus (33s), tyydytys (33s), tärkeys (33s), uhma (33s), uho (33s), uhrautuvuus (33s), urheus (33s), vajavuus (33s), valoisuus (33s), vilkkaus (33s), vilu (33s), virkeys (33s), ylemmyys (33s), ylenkatse (33s), äreys (33s).

# Appendix B

**Table B1.** The visual scale and main instructions provided for the person of experimental evaluation about the Self Assessment Manikin (SAM) test based on the instructions of Bradley and Lang (1994; 1999a). For each of the total of 211 concepts the person had to provide a rating value on a five-point visual scale for three affectivity dimensions of pleasure, arousal and dominance. During the rating task for each word the three SAM scales were shown separately one at time. The person answered by selecting one of the five pictures of the scale and the selected answers were recorded as numbers in the order of rising affectivity by the values 1, 2, 3, 4 and 5 (in the order shown here for pictures from left to right).

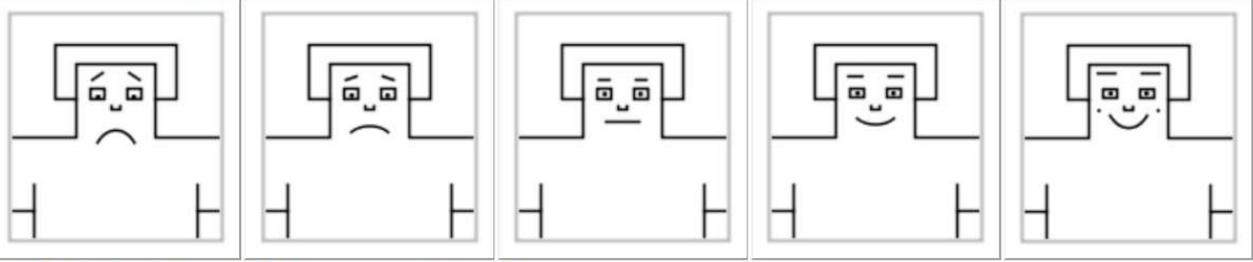

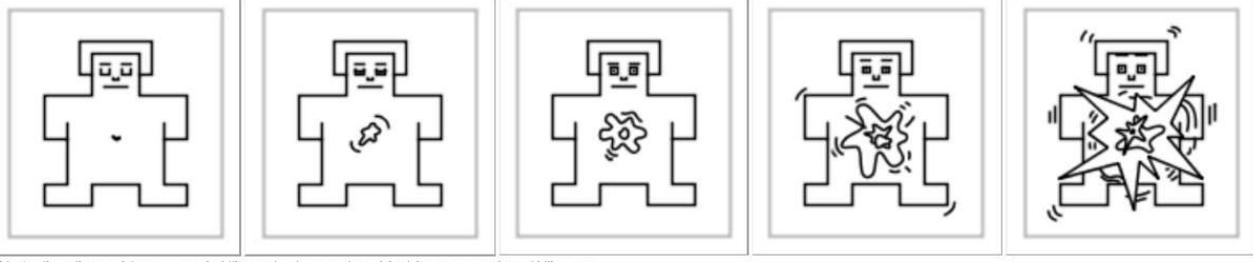

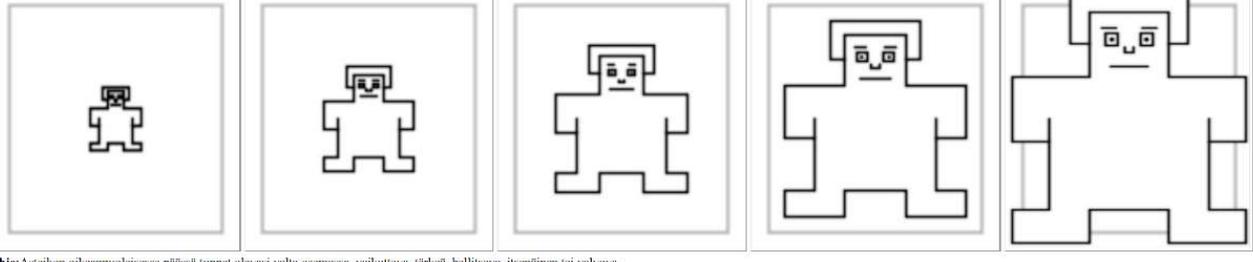

# Appendix C

**Table C1.** Comparison of Pearson product-moment correlation coefficient measures for affective rating measures between the pairs of the dimensions pleasure, arousal and dominance for several population subgroup categorizations in respect to 195 emotional adjectives and 16 pregnancy-related nouns.

| Group | Comparison of correlation coefficient measures concerning 195 emotional adjectives | | | Comparison of correlation coefficient measures concerning 16 preganancy-related nouns | | |
|---|---|---|---|---|---|---|
| | *pleasure & arousal* | *pleasure & dominance* | *arousal & dominance* | *pleasure & arousal* | *pleasure & dominance* | *arousal & dominance* |
| All (n=35) | -0.18 (p < 0.001) | 0.75 (p < 0.001) | -0.16 (p < 0.001) | -0.25 (p < 0.001) | 0.67 (p < 0.001) | -0.22 (p < 0.001) |
| Women (n=21) | -0.19 (p < 0.001) | 0.79 (p < 0.001) | -0.15 (p < 0.001) | -0.25 (p < 0.001) | 0.74 (p < 0.001) | -0.2 (p < 0.001) |
| Men (n=14) | -0.16 (p < 0.001) | 0.69 (p < 0.001) | -0.17 (p < 0.001) | -0.26 (p < 0.001) | 0.55 (p < 0.001) | -0.24 (p < 0.001) |
| Women without children (n=13) | -0.19 (p < 0.001) | 0.8 (p > 0.05) | -0.14 (p > 0.05) | -0.31 (p < 0.001) | 0.73 (p < 0.001) | -0.28 (p < 0.001) |
| Women with children (n=8) | -0.19 (p < 0.001) | 0.77 (p < 0.001) | -0.16 (p < 0.001) | -0.13 (p > 0.05) | 0.75 (p < 0.001) | -0.09 (p > 0.05) |
| Men without children (n=13) | -0.17 (p < 0.001) | 0.68 (p < 0.001) | -0.17 (p < 0.001) | -0.27 (p < 0.001) | 0.53 (p < 0.001) | -0.23 (p < 0.001) |

**Table C2.** The distribution of 6825 rating answers (shown here in percent with the accuracy rounded to two digits) of 35 persons in respect to 125 alternative answer categories based on three dimensions of measures for 195 emotional adjectives. For each five levels of dominance a two-dimensional table is shown for pleasure and arousal.

**Table C3.** Comparison of Pearson product-moment correlation coefficient measures for average affective rating measures between several collections for 195 emotional adjectives and 16 pregnancy-related nouns.

| Compared collections (full population samples) | Correlation coefficients for emotional words | | Correlation coefficients for pregnancy-related words | |
|---|---|---|---|---|
| | *pleasure (adjusted to the range [-2,2])* | *arousal (adjusted to the range [-2,2])* | *pleasure (adjusted to the range [-2,2])* | *arousal (adjusted to the range [-2,2])* |
| Current research data (n=35) vs. Söderholm et al. (2013) | 0.772 (55 words) (p < 0.001) | 0.781 (55 words) (p < 0.001) | 0.880 (3 words) (p > 0.05) | -0.305 (3 words) (p > 0.05) |
| Current research data (n=35) vs. Eilola & Havelka (2010) | 0.820 (35 words) (p < 0.001) | 0.072 (35 words) (p > 0.05) | -0.726 (3 words) (p > 0.05) | 0.655 (3 words) (p > 0.05) |

**Table C4.** Comparison of Pearson product-moment correlation coefficient measures for average affective rating measures between several collections for the six basic emotion categories proposed by Ekman (1972).

| Compared collections (full population samples) | Correlation coefficients for the six basic emotion categories proposed by Ekman (1972) | | |
|---|---|---|---|
| | *pleasure (adjusted to the range [-2,2])* | *arousal (adjusted to the range [-2,2])* | *dominance (adjusted to the range [-2,2])* |
| Current research data (n=35) vs. Söderholm et al. (2013) | Not defined (2 words) | Not defined (2 words) | N/A |
| Current research data (n=35) vs. Eilola & Havelka (2010) | 0.996 (3 words) (p = 0.05) | 0.315 (3 words) (p > 0.05) | N/A |
| Current research data (n=35) vs. adjective lemmas of Warriner et al. (2013) | 0.998 (6 words) (p < 0.001) | 0.724 (6 words) (p > 0.05) | 0.854 (6 words) (p < 0.04) |
| Current research data (n=35) vs. noun lemmas of Warriner et al. (2013) | 0.986 (6 words) (p < 0.001) | 0.707 (6 words) (p > 0.05) | 0.929 (6 words) (p < 0.008) |

**Evaluation of belonging to a subgroup:**

Concerning subgroups gender, gender-parental, rating-daytime and rating-duration it appeared that ANOVA tests indicated that belonging to a subgroup had same kinds of statistically significant main effect patterns on affective rating dimensions for all these four subgroup categorizations. An ANOVA test indicated that the main effect of belonging to a gender subgroup on the combined set of all three affective rating dimensions was statistically significant, $F(3, 6821) = 4.689$, $p < 0.003$. In a same way ANOVA tests indicated that the main effect of belonging to a subgroup on the combined set of all three affective rating dimensions was statistically significant for the subgroups gender-parental, $F(3, 6626) = 5.8986$, $p < 0.001$, rating-daytime, $F(3, 6821) = 18.654$, $p < 0.001$, and rating-duration, $F(3, 6821) = 22.75$, $p < 0.001$.

**Evaluation of gender subgroups:**

ANOVA tests indicated that the main effect of belonging to a gender subgroup on arousal dimension was statistically significant, $F(1, 6823) = 7.8513$, $p < 0.006$, but the main effect of belonging to a gender subgroup on pleasure dimension was statistically non-significant, $F(1, 6823) = 2.9103$, $p > 0.05$. and the main effect of belonging to a gender subgroup on dominance dimension was statistically non-significant, $F(1, 6823) = 3.7382$, $p > 0.05$.

**Evaluation of gender-parental subgroups:**

ANOVA tests indicated that the main effect of belonging to a gender-parental subgroup on arousal dimension was statistically significant, $F(1, 6628) = 10.94$, $p < 0.001$, and the main effect of belonging to a gender-parental subgroup on dominance dimension was statistically significant, $F(1, 6628) = 4.1632$, $p < 0.05$, but the main effect of belonging to a gender-parental subgroup on pleasure dimension was statistically non-significant, $F(1, 6628) = 1.564$, $p > 0.05$,

**Evaluation of rating-daytime subgroups:**

ANOVA tests indicated that the main effect of belonging to a rating-daytime subgroup on arousal dimension was statistically significant, $F(1, 6823) = 40.743$, $p < 0.001$, and the main effect of belonging to a rating-daytime subgroup on dominance dimension was statistically significant, $F(1, 6823) = 7.602$, $p < 0.006$, but the main effect of belonging to a rating-daytime subgroup on pleasure dimension was statistically non-significant, $F(1, 6823) = 1.3419$, $p > 0.05$,

**Evaluation of rating-duration subgroups:**

ANOVA tests indicated that the main effect of belonging to a rating-duration subgroup on arousal dimension was statistically significant, $F(1, 6823) = 32.178$, $p < 0.001$. and the main effect of belonging to a rating-duration subgroup on dominance duration was statistically significant, $F(1, 6628) = 11.867$, $p < 0.001$, but the main effect of belonging to a rating-duration subgroup on pleasure dimension was statistically non-significant, $F(1, 6628) = 0.362$, $p > 0.05$.

**Table C5.** Average and standard deviation for the affective ratings in respect to the time of the day of answering (n=35).

| Time of the day when the person started giving affective rating answers | pleasure | | arousal | | dominance | |
|---|---|---|---|---|---|---|
| Group | average | stand. dev. | average | stand. dev. | average | stand. dev. |
| Early, between 07:45 and 12:28 (n=11) | -0.08 | 1.54 | -0.2 | 1.31 | -0.24 | 1.4 |
| Middle, between 12:28 and 18:43 (n=12) | 0 | 1.39 | -0.26 | 1.31 | -0.12 | 1.22 |
| Late, between 18:43 and 23:38 (n=12) | -0.03 | 1.46 | 0.04 | 1.33 | -0.13 | 1.29 |

**Table C6.** Average and standard deviation for the affective ratings in respect to the average duration of answering (n=35).

| Average duration of three affective rating answers given for a word by the person | pleasure | | arousal | | dominance | |
|---|---|---|---|---|---|---|
| Group | average | stand. dev. | average | stand. dev. | average | stand. dev. |
| Short, 8.246 – 12.406 seconds (n=11) | -0.05 | 1.57 | 0 | 1.35 | -0.24 | 1.37 |
| Medium, 12.406 – 16.837 seconds (n=12) | 0.02 | 1.38 | -0.18 | 1.29 | -0.16 | 1.28 |
| Long, 16.837 – 38.303 seconds (n=12) | -0.08 | 1.43 | -0.22 | 1.32 | -0.1 | 1.27 |

**Evaluation of the difference of affective ratings in respect to gender and age:**

When aiming to replicate a previously noted pattern that typically men provide higher affective ratings than women and younger people higher ratings than older people in all three dimensions (Warriner et al. (2013), our ANOVA tests indicated that for 195 emotional adjectives the combined set of all three affective rating dimensions in respect to the gender subgroups "women" "men" differed statistically significantly ($F(3, 2726) = 214.37$, $p < 0.001$).

Similarly for 195 emotional adjectives the combined set of all three affective rating dimensions differed statistically significantly in respect to the gender-parental subgroups "women with children" and "women without children", $F(3, 1556) = 186.87$, $p < 0.001$, "women with children" and "men without children", $F(3, 1556) = 112.4$, $p < 0.001$, as well as "women without children" and "men without children", $F(3, 2531) = 198.01$, $p < 0.001$.

In addition our ANOVA tests indicated that for 195 emotional adjectives affective ratings between gender subgroups women and men differed significantly for pleasure, $F(1, 2728) = 2943$, $p < 0.001$, arousal, $F(1, 2728) = 93.8$, $p < 0.001$, and dominance, $F(1, 2728) = 939.5$, $p < 0.001$.

Furthermore our ANOVA tests indicated that for 195 emotional adjectives affective ratings between gender-parental subgroups "women with children" and "women without children" differed significantly for pleasure, $F(1, 1558) = 3685$, $p < 0.001$, and arousal, $F(1, 1558) = 95.49$, $p < 0.001$, and non-significantly for dominance, $F(1, 1558) = 0.015$, $p > 0.05$. Similarly affective ratings between gender-parental subgroups "women with children" and "men without children" differed significantly for pleasure, $F(1, 1558) = 1432$, $p < 0.001$, arousal, $F(1, 1558) = 40.8$, $p < 0.001$, and dominance, $F(1, 1558) = 502.5$, $p < 0.05$. In a same way affective ratings between gender-parental subgroups "women without children" and "men without children" differed significantly for pleasure, $F(1, 2533) = 2775$, p

< 0.001, and arousal, F(1, 2533) = 113.3, p < 0.001, and non-significantly for dominance, F(1, 2533) = 2.055, p > 0.05 (see Table 2 and Appendix C).

**Evaluation of the difference of arousal ratings in respect to pregnancy and motherhood:**
In an attempt to replicate at least indirectly the previous results of Rosebrock et al. (2015) we carried out ANOVA tests that indicated that the arousal ratings of the subgroup "women with children" and the subgroup "women without children" differed statistically significantly for 195 emotional adjectives, F(1, 1558) = 95.49, p < 0.001, whereas they did not differ statistically significantly for 16 pregnancy-related nouns, F(1, 126) = 0.126, p > 0.05.

**Evaluation of the time usage for giving the affective rating of pleasure:**
We evaluated the dependence between the affective ratings of the pleasure dimension and the time usage for giving the affective rating of pleasure which was measured with the accuracy level of a second. For 121 of our 195 emotional adjectives we found a match with the 50 000 highest-frequent words of Parole corpus of Finnish written language (Parole corpus 2017). In respect to the pleasure ratings of full population sample (n=35) we classified 60 highest-frequent words of these 121 words as positive (pleasure rating above 0.667), neutral (pleasure rating in the range [-0.667,0.667]) or negative (pleasure rating below 0.667).

With an effort to balance the frequency effect (i.e. a person typically responds faster to high-frequency than low-frequency stimulus words of a language (Duyck et al., 2008)) we selected a subsample of ten positive words and ten negative high-frequency words so that the average of their frequency ranking positions was 7515.7 for positive words and 7488.4 for negative words (positive words included amazing (mahtava), wonderful (loistava), proud (ylpeä), excited (innostunut), safe (turvallinen), lovely (ihana), committed (sitoutunut), whole (kokonainen), kind (kiltti) and superb (super), and negative words included longing (ikävä), empty (tyhjä), tired (väsynyt), disappointed (pettynyt), horrible (hirveä), concerned (huolestunut), lonely (yksinäinen), negative (kielteinen), guilty (syyllinen) and miserable_II (onneton)).

We measured the time used for giving the affective rating answer for the dimension of pleasure immediately after the word was shown for the first time. It turned out that below a cut-off level of 10 seconds the response time in seconds for the set of positive words (M = 2.95, SD = 1.46) and for the set of negative words (M = 3.02, SD = 1.61) did not differ significantly, t(623.65) = -0.54, p > 0.05

**Table C7.** For all five clusters of each gender-parental subgroup the measures of precision, recall and F1 to indicate the amount of matching of clustering of subgroups.

| When comparing subgroup "Women with children" (n=8) to subgroup "Women without children" (n=13) | | | | | | |
|---|---|---|---|---|---|---|
| *measure* | *A* | *B* | *C* | *D* | *E* | *average* |
| precision | 0.64 | 0.81 | 0.89 | 0.61 | 0.71 | 0.73 |
| recall | 0.82 | 0.66 | 0.76 | 0.67 | 0.77 | 0.73 |
| F1 score | 0.72 | 0.72 | 0.82 | 0.64 | 0.74 | 0.73 |
| | | | | | | |
| When comparing subgroup "Men without children" (n=13) to subgroup "Women without children" (n=13) | | | | | | |
| *measure* | *A* | *B* | *C* | *D* | *E* | *average* |
| precision | 0.68 | 0.88 | 0.91 | 0.52 | 0.55 | 0.71 |
| recall | 0.85 | 0.44 | 0.82 | 0.52 | 0.85 | 0.70 |
| F1 score | 0.76 | 0.58 | 0.86 | 0.52 | 0.67 | 0.70 |
| | | | | | | |
| When comparing subgroup "Men without children" (n=13) to subgroup "Women with children" (n=8) | | | | | | |
| *measure* | *A* | *B* | *C* | *D* | *E* | *average* |
| precision | 0.73 | 0.75 | 0.77 | 0.64 | 0.60 | 0.70 |
| recall | 0.71 | 0.46 | 0.81 | 0.59 | 0.86 | 0.69 |
| F1 score | 0.72 | 0.57 | 0.79 | 0.61 | 0.71 | 0.69 |

**Table C8.** Comparison of average affective rating measures between the collections of the current research data and Warriner et al. (2013) for 16 pregnancy-related nouns that we considered semantically most close.

| Current research data of 16 pregnancy-related nouns (n=35), full population samples | | | | Collection of Warriner et al. (2013), full population samples | | | | Cosine similarity measure between average vectors of words |
|---|---|---|---|---|---|---|---|---|
| *word* | *pleasure (adjusted to the range [-2,2])* | *arousal (adjusted to the range [-2,2])* | *dominance (adjusted to the range [-2,2])* | *word* | *pleasure (adjusted to the range [-2,2])* | *arousal (adjusted to the range [-2,2])* | *dominance (adjusted to the range [-2,2])* | |
| intimate relationship | 1.11 | -0.26 | 0.66 | couple | 1.06 | -0.03 | 0.31 | 0.95 |
| motherhood | 0.89 | -0.06 | 0.2 | motherhood | 0.9 | 0.26 | 0.5 | 0.91 |
| fatherhood | 1 | -0.51 | 0.46 | fatherhood | 0.89 | -0.22 | 0.31 | 0.98 |
| infant | 0.94 | -0.4 | 0.06 | baby | 0.84 | -0.02 | -0.03 | 0.92 |
| fetus | 0.23 | -0.09 | -0.43 | fetus | -0.13 | -0.38 | -0.2 | 0.41 |
| pregnancy | 0.6 | 0.14 | -0.23 | pregnancy | -0.11 | 0 | -0.22 | -0.1 |
| giving birth | 0 | 0.71 | -0.77 | birth | 0.76 | 0.38 | 0 | 0.3 |
| breastfeeding | 0.74 | -0.54 | 0.09 | nursing | 0.7 | -0.72 | 0.03 | 0.98 |
| baby colic | -1.26 | 0.63 | -1.17 | colic | -1.03 | -1.06 | -1.03 | 0.56 |
| miscarriage | -1.83 | 0.97 | -1.86 | miscarriage | -1.26 | 0.35 | -1.34 | 0.99 |
| abortion | -1.03 | 0.31 | -0.66 | abortion | -1.21 | 0.22 | -0.14 | 0.9 |
| preemie | -0.69 | 0.29 | -1.11 | premature | -0.45 | -0.19 | -0.3 | 0.77 |
| childlessness | -1.2 | -0.43 | -1.11 | childless | -0.72 | -0.86 | 0.4 | 0.39 |
| sexuality | 1.46 | 0.4 | 0.51 | sexuality | 0.48 | 1.07 | 0.87 | 0.67 |
| sole custody of child | -0.54 | -0.11 | -0.49 | custody | -0.13 | -0.43 | 0.27 | -0.04 |
| artificial fertilization | 0.2 | -0.4 | -0.34 | fertilize | 0.23 | -0.37 | 0.76 | -0.13 |

**Table C9.** Comparison of Pearson product-moment correlation coefficient measures for average affective rating measures between the collections of the current research data (n=35) and Warriner et al. (2013) for 16 pregnancy-related nouns that we considered semantically most close.

| Compared collections (full population samples) | Correlation coefficients for pregnancy-related words | | |
|---|---|---|---|
| | *pleasure (adjusted to the range [-2,2])* | *arousal (adjusted to the range [-2,2])* | *dominance (adjusted to the range [-2,2])* |
| Current research data (n=35) vs. Warriner et al. (2013) | 0.900 (16 words) (p < 0.001) | 0.417 (16 words) (p > 0.05) | 0.711 (16 words) (p < 0.002) |

**Table C10.** Comparison of Pearson product-moment correlation coefficient measures and cosine similarity measures between the collections of gender-parental subgroups and gender subgroups of Warriner et al. (2013) for Finnish words "sikiö", "raskaus" and "synnytys" (translated in English as "fetus", "pregnancy" and "giving birth" respectively).

| Compared collections (partial segments of full population samples) | Correlation coefficients for the words sikiö, raskaus and synnytys | | | Cosine similarity measure between average vectors of words | | |
|---|---|---|---|---|---|---|
| | *pleasure* | *arousal* | *dominance* | *sikio* | *raskaus* | *synnytys* |
| Women without children (n=13) vs. Women responses of Warriner et al. (2013) | -0.221 (3 words) (p > 0.05) | 0.966 (3 words) (p > 0.05) | -0.254 (3 words) (p > 0.05) | 0.063 | 0.309 | 0.007 |
| Women with children (n=8) vs. Women responses of Warriner et al. (2013) | -0.866 (3 words) (p > 0.05) | 0.999 (3 words) (p < 0.05) | 0.500 (3 words) (p > 0.05) | -0.063 | 0.722 | 0.530 |
| Men without children (n=13) vs. Men responses of Warriner et al. (2013) | -0.593 (3 words) (p > 0.05) | 0.332 (3 words) (p > 0.05) | -0.963 (3 words) (p > 0.05) | 0.126 | -0.575 | -0.031 |

## Appendix D

**Table D1.** Some general transformation vectors in respect to various subgroup categories.

| Subgroup category | Tranformation vectors |
|---|---|
| Gender subgroups | $T_{women→men}$ = (+0.06; +0.09; +0.06) |
| Gender-parental subgroups | $T_{women\_with\_children→women\_without\_children}$ = (0; +0.15; +0.03) <br> $T_{women\_with\_children→men\_without\_children}$ = (+0.05; +0.16; +0.08) <br> $T_{women\_without\_children→men\_without\_children}$ = (+0.05; +0.01; +0.05) |
| Rating-daytime subgroups | $T_{early→middle}$ = (+0.08; -0.06; +0.12) <br> $T_{early→late}$ = (+0.05; +0.24; +0.11) <br> $T_{middle→late}$ = (-0.03; +0.30; -0.01) |
| Rating-duration subgroups | $T_{short→medium}$ = (+0.07; -0.18; +0.08) <br> $T_{short→long}$ = (-0.03; -0.22; +0.14) <br> $T_{medium→long}$ = (-0.10; -0.04; +0.06) |

**Table D2.** In respect to the word-specific transformation model (emotional adjectives) some of the word-specific transformation vectors for gender-parental subgroups shown in a descending order of the distance between a pair of gender-parental subgroups.

| Transformation from subgroup Women without children to subgroup Women with children | | Transformation from subgroup Women without children to subgroup Men without children | | Transformation from subgroup Women with children to subgroup Men without children | |
|---|---|---|---|---|---|
| word-specific transformaton vector | distance between subgroups | word-specific transformation vector | distance between subgroups | word-specific transformaton vector | distance between subgroups |
| $WT_{women\_without\_children→women\_with\_children}$("euphoric") = (-0.27; -1.49; -0.85) | 1.73 | $WT_{women\_without\_children→men\_without\_children}$("lost") = (0.54; -1.38; 1) | 1.79 | $WT_{women\_with\_children→men\_without\_children}$("naked") = (0.92; 1.51; 1.01) | 2.04 |
| $WT_{women\_without\_children→women\_with\_children}$("naked") = (-0.15; -1.51; 0.22) | 1.53 | $WT_{women\_without\_children→men\_without\_children}$("aggressive") = (0.54; -1.15; 0.92) | 1.57 | ("euphoric") = (-0.27; 1.57; 0.46) | 1.66 |
| $WT_{women\_without\_children→women\_with\_children}$("terrible") = (0.19; 0.57; -1.03) | 1.19 | $WT_{women\_without\_children→men\_without\_children}$("burned out") = (0.08; 1.54; -0.08) | 1.54 | ("wow!") = (-0.83; -0.88; -0.91) | 1.51 |
| $WT_{women\_without\_children→women\_with\_children}$("aggressive") = (0.21; -0.71; 0.92) | 1.18 | $WT_{women\_without\_children→men\_without\_children}$("special") = (-0.92; 0.69; -1) | 1.53 | ("terrible") = (0.58; -0.88; 1.03) | 1.47 |
| $WT_{women\_without\_children→women\_with\_children}$("uncomfortable") = (0.26; -1.14; 0.13) | 1.18 | $WT_{women\_without\_children→men\_without\_children}$("naked") = (0.77; 0; 1.23) | 1.45 | ("appreciative") = (-1.13; 0.62; -0.54) | 1.4 |
| $WT_{women\_without\_children→women\_with\_children}$("fearless") = (0.31; 0.2; 1.04) | 1.1 | $WT_{women\_without\_children→men\_without\_children}$("desperate") = (0.38; -1.15; 0.77) | 1.44 | ("superb") = (-0.67; 0.29; -1.09) | 1.31 |
| $WT_{women\_without\_children→women\_with\_children}$("shared sense of shame") = (-0.15; -0.99; -0.44) | 1.1 | $WT_{women\_without\_children→men\_without\_children}$("abandoned") = (0.62; -1; 0.54) | 1.29 | ("shocked") = (0.64; -0.78; 0.78) | 1.28 |
| $WT_{women\_without\_children→women\_with\_children}$("awkward") = (-0.15; -0.94; 0.45) | 1.06 | $WT_{women\_without\_children→men\_without\_children}$("energetic") = (-0.31; -0.08; -1.23) | 1.27 | ("ecstatic") = (-0.83; 0.45; -0.83) | 1.25 |
| $WT_{women\_without\_children→women\_with\_children}$("self-pitying") = (0.42; -0.21; -0.92) | 1.04 | $WT_{women\_without\_children→men\_without\_children}$("disappointed") = (0.15; 1; 0.77) | 1.27 | ("blessed") = (-0.51; 1.12; -0.21) | 1.24 |
| $WT_{women\_without\_children→women\_...}$ | 1.03 | $WT_{women\_without\_children→men\_without\_children}$ | 1.26 | ("energetic") = (-0.63; | 1.24 |
| $WT_{women\_without\_children}$("tense_l") = (-0.18; -1.01; 0.05) | | $WT_{men\_without\_children}$("shocked") = (0.54; -0.54; 1) | | 0.54; -0.92) | |
| $WT_{women\_without\_children \rightarrow women\_with\_children}$("empty") = (-0.19; 0.88; -0.47) | 1.02 | $WT_{women\_without\_children \rightarrow men\_without\_children}$("alarmed") = (0.69; -0.62; 0.85) | 1.25 | ("special") = (-0.36; 0.78; -0.88) | 1.22 |
| $WT_{women\_without\_children \rightarrow women\_with\_children}$("ecstatic") = (0.06; -0.76; 0.67) | 1.02 | $WT_{women\_without\_children \rightarrow men\_without\_children}$("powerless") = (0.85; 0.54; 0.69) | 1.22 | ("depressed") = (0.54; 0.87; 0.62) | 1.19 |
| $WT_{women\_without\_children \rightarrow women\_with\_children}$("lost") = (0.31; -0.88; 0.41) | 1.02 | $WT_{women\_without\_children \rightarrow men\_without\_children}$("skeptical") = (0.77; -0.54; 0.77) | 1.21 | ("small") = (-0.74; 0.26; -0.89) | 1.19 |
| $WT_{women\_without\_children \rightarrow women\_with\_children}$("small") = (0.59; -0.49; 0.66) | 1.01 | $WT_{women\_without\_children \rightarrow men\_without\_children}$("appreciative") = (-1; 0.38; -0.46) | 1.17 | ("upset") = (0.49; -0.46; 0.96) | 1.17 |
| $WT_{women\_without\_children \rightarrow women\_with\_children}$("skeptical") = (0.15; -0.91; 0.38) | 1 | $WT_{women\_without\_children \rightarrow men\_without\_children}$("anxious") = (0.77; -0.85; 0.08) | 1.15 | ("nervous") = (0.86; 0.44; 0.66) | 1.17 |
| $WT_{women\_without\_children \rightarrow women\_with\_children}$("nervous") = (-0.39; -0.9; -0.05) | 0.99 | $WT_{women\_without\_children \rightarrow men\_without\_children}$("excited") = (-0.62; -0.85; -0.46) | 1.14 | ("fantastic") = (-0.32; 1.1; -0.23) | 1.16 |
| $WT_{women\_without\_children \rightarrow women\_with\_children}$("restless") = (0.18; -0.86; -0.43) | 0.98 | $WT_{women\_without\_children \rightarrow men\_without\_children}$("manic") = (0.38; -1; 0.38) | 1.14 | ("numb") = (0.81; -0.08; 0.83) | 1.16 |
| $WT_{women\_without\_children \rightarrow women\_with\_children}$("helpless") = (0.32; -0.88; 0.19) | 0.96 | $WT_{women\_without\_children \rightarrow men\_without\_children}$("sad") = (0.15; 0.92; 0.62) | 1.12 | ("gloating") = (0.61; 0.94; 0.26) | 1.15 |
| $WT_{women\_without\_children \rightarrow women\_with\_children}$("alert") = (0; -0.94; 0.06) | 0.94 | $WT_{women\_without\_children \rightarrow men\_without\_children}$("lonely") = (0.62; 0.62; 0.69) | 1.11 | ("disoriented") = (0.96; -0.28; 0.53) | 1.13 |
| $WT_{women\_without\_children \rightarrow women\_with\_children}$("privileged") = (0.36; -0.69; -0.53) | 0.94 | $WT_{women\_without\_children \rightarrow men\_without\_children}$("upset") = (0.38; -0.69; 0.77) | 1.1 | ("harmonious") = (-0.22; 1.06; 0.14) | 1.09 |
| $WT_{women\_without\_children \rightarrow women\_with\_children}$("powerless") = (0.27; 0.38; 0.8) | 0.93 | $WT_{women\_without\_children \rightarrow men\_without\_children}$("cohesive") = (-0.46; 0.85; -0.54) | 1.1 | ("productive") = (-0.64; -0.79; -0.38) | 1.09 |
| $WT_{women\_without\_children \rightarrow women\_with\_children}$("anxious") = (0.35; -0.85; 0.06) | 0.92 | $WT_{women\_without\_children \rightarrow men\_without\_children}$("blessed") = (0; 1; 0.46) | 1.1 | ("astonished") = (-0.22; 1.01; 0.33) | 1.08 |
| $WT_{women\_without\_children \rightarrow women\_with\_children}$("unique") = (-0.22; -0.82; 0.31) | 0.9 | $WT_{women\_without\_children \rightarrow men\_without\_children}$("restless") = (0.77; -0.77; -0.08) | 1.09 | ("amused") = (-0.25; 0.88; -0.58) | 1.08 |
| $WT_{women\_without\_children \rightarrow women\_with\_children}$("numb") = (-0.27; 0.15; -0.83) | 0.88 | $WT_{women\_without\_children \rightarrow men\_without\_children}$("nurturing") = (-0.08; 1.08; 0.15) | 1.09 | ("longing") = (0.71; 0.17; 0.79) | 1.08 |
| $WT_{women\_without\_children \rightarrow women\_with\_children}$("sad") = (0.24; 0.73; 0.43) | 0.88 | $WT_{women\_without\_children \rightarrow men\_without\_children}$("drained") = (0.62; -0.46; 0.77) | 1.09 | ("unsafe") = (0.7; 0.35; 0.72) | 1.06 |
| $WT_{women\_without\_children \rightarrow women\_with\_children}$("astonished") = (0.14; -0.86; 0.13) | 0.88 | $WT_{women\_without\_children \rightarrow men\_without\_children}$("insecure") = (0.62; -0.77; 0.38) | 1.06 | ("interested") = (-0.3; 0.91; 0.44) | 1.06 |
| $WT_{women\_without\_children \rightarrow women\_with\_children}$("admiring") = (-0.11; -0.75; -0.44) | 0.88 | $WT_{women\_without\_children \rightarrow men\_without\_children}$("hopeless") = (0.38; 0.62; 0.77) | 1.06 | ("cynical") = (0.73; 0.57; 0.49) | 1.05 |
| $WT_{women\_without\_children \rightarrow women\_{...}}$ | 0.87 | $WT_{women\_without\_children \rightarrow ...}$ | 1.05 | ("privileged") = (-0.66; | 1.04 |
| $WT_{women\_with\_children}$("disappointed") = (0.31; 0.82; 0.03) | | $WT_{men\_without\_children}$("irresolute") = (0.31; -1; 0.08) | | (0.69; -0.39) | |
| $WT_{women\_without\_children \to women\_with\_children}$("distrustful") = (0.18; -0.84; 0.17) | 0.87 | $WT_{women\_without\_children \to men\_without\_children}$("concerned") = (0.38; -0.31; 0.92) | 1.05 | ("drained") = (0.64; -0.35; 0.72) | 1.03 |
| $WT_{women\_without\_children \to women\_with\_children}$("confident") = (-0.69; -0.03; -0.52) | 0.87 | $WT_{women\_without\_children \to men\_without\_children}$("wow!") = (-0.69; -0.62; -0.46) | 1.03 | ("unique") = (-0.01; 0.82; -0.62) | 1.02 |
| $WT_{women\_without\_children \to women\_with\_children}$("blessed") = (0.51; -0.12; 0.67) | 0.85 | $WT_{women\_without\_children \to men\_without\_children}$("indifferent") = (0.62; 0.77; 0.15) | 1 | ("fearful") = (0.37; -0.94; -0.13) | 1.02 |
| $WT_{women\_without\_children \to women\_with\_children}$("reassured") = (-0.71; -0.42; 0.16) | 0.84 | $WT_{women\_without\_children \to men\_without\_children}$("whole") = (-0.31; 0.85; -0.38) | 0.98 | ("burned out") = (0.38; 0.91; 0.18) | 1.01 |
| $WT_{women\_without\_children \to women\_with\_children}$("hurt") = (0.3; 0.77; 0.16) | 0.84 | $WT_{women\_without\_children \to men\_without\_children}$("miserable_II") = (0.77; 0.46; 0.38) | 0.98 | ("loved") = (-0.29; 0.22; -0.93) | 1 |
| $WT_{women\_without\_children \to women\_with\_children}$("remorseful") = (0.48; -0.63; 0.28) | 0.84 | $WT_{women\_without\_children \to men\_without\_children}$("stressed") = (0.31; -0.62; 0.69) | 0.98 | ("judgmental") = (0.38; 0.15; 0.91) | 1 |
| $WT_{women\_without\_children \to women\_with\_children}$("pitying") = (0.46; -0.51; 0.47) | 0.83 | $WT_{women\_without\_children \to men\_without\_children}$("privileged") = (-0.31; 0; -0.92) | 0.97 | ("fearless") = (-0.38; 0.64; -0.65) | 1 |
| $WT_{women\_without\_children \to women\_with\_children}$("mistreated") = (0.02; 0.82; 0.12) | 0.83 | $WT_{women\_without\_children \to men\_without\_children}$("ashamed") = (0.38; -0.85; 0.23) | 0.96 | ("admiring") = (-0.05; 0.98; -0.1) | 0.99 |
| $WT_{women\_without\_children \to women\_with\_children}$("optimistic") = (-0.66; -0.47; -0.1) | 0.82 | $WT_{women\_without\_children \to men\_without\_children}$("interested") = (-0.23; 0.92; -0.08) | 0.95 | ("miserable_I") = (-0.34; 0.17; 0.98) | 0.98 |
| $WT_{women\_without\_children \to women\_with\_children}$("nurturing") = (0.13; 0.57; 0.57) | 0.81 | $WT_{women\_without\_children \to men\_without\_children}$("tense_I") = (0.38; -0.62; 0.62) | 0.95 | ("alarmed") = (0.67; -0.4; 0.58) | 0.97 |
| $WT_{women\_without\_children \to women\_with\_children}$("relieved") = (-0.56; -0.58; -0.12) | 0.81 | $WT_{women\_without\_children \to men\_without\_children}$("contemptuous") = (0.54; -0.46; 0.62) | 0.94 | ("empty") = (0.42; -0.73; 0.47) | 0.97 |
| $WT_{women\_without\_children \to women\_with\_children}$("determined to win") = (0.08; -0.8; -0.03) | 0.8 | $WT_{women\_without\_children \to men\_without\_children}$("fearless") = (-0.08; 0.85; 0.38) | 0.93 | ("helpless") = (-0.09; 0.81; 0.5) | 0.95 |

**Table D3.** In respect to the word-specific transformation model (pregnancy-related nouns) some of the word-specific transformation vectors for gender-parental subgroups shown in a descending order of the distance between a pair of gender-parental subgroups.

| Transformation from subgroup Women without children to subgroup Women with children | | Transformation from subgroup Women without children to subgroup Men without children | | Transformation from subgroup Women with children to subgroup Men without children | |
|---|---|---|---|---|---|
| word-specific transformaton vector | distance between subgroups | word-specific transformaton vector | distance between subgroups | word-specific transformaton vector | distance between subgroups |
| $WT_{women\_without\_children \to women\_with\_children}$("motherhood") = (1.13; -0.23; 1.36) | 1.78 | $WT_{women\_without\_children \to men\_without\_children}$("childlessness") = (0.46; 0.16; 0.93) | 1.05 | $WT_{women\_with\_children \to men\_without\_children}$("motherhood") = (-1.13; -0.31; -1.13) | 1.63 |
| $WT_{women\_without\_children \to women\_with\_children}$("fetus") = (1.31; -0.31; 0.54) | 1.45 | $WT_{women\_without\_children \to men\_without\_children}$("fetus") = (0.54; -0.85; -0.15) | 1.02 | $WT_{women\_with\_children \to men\_without\_children}$("infant") = (-1.01; -0.44; -1.11) | 1.56 |
| $WT_{women\_without\_children \to women\_with\_children}$("giving birth") = (0.64; 0.36; 1.15) | 1.36 | $WT_{women\_without\_children \to men\_without\_children}$("pregnancy") = (0.07; -0.93; -0.08) | 0.94 | $WT_{women\_with\_children \to men\_without\_children}$("childlessness") = (0.65; -0.08; 1.17) | 1.34 |
| $WT_{women\_without\_children \to women\_with\_children}$("infant") = (0.78; -0.1; 1.11) | 1.36 | $WT_{women\_without\_children \to men\_without\_children}$("giving birth") = (0.54; -0.46; 0.3) | 0.77 | $WT_{women\_with\_children \to men\_without\_children}$("giving birth") = (-0.1; -0.82; -0.85) | 1.19 |
| $WT_{women\_without\_children \to women\_with\_children}$("artificial fertilization") = (0.71; -0.92; 0.18) | 1.18 | $WT_{women\_without\_children \to men\_without\_children}$("baby colic") = (0.38; -0.38; 0.54) | 0.76 | $WT_{women\_with\_children \to men\_without\_children}$("fetus") = (-0.77; -0.54; -0.69) | 1.17 |
| $WT_{women\_without\_children \to women\_with\_children}$("fatherhood") = (0.69; 0.29; 0.69) | 1.02 | $WT_{women\_without\_children \to men\_without\_children}$("preemie") = (0.23; -0.62; 0.23) | 0.7 | $WT_{women\_with\_children \to men\_without\_children}$("pregnancy") = (-0.67; -0.64; -0.64) | 1.13 |
| $WT_{women\_without\_children \to women\_with\_children}$("sole custody of child") = (0.77; -0.3; 0.54) | 0.99 | $WT_{women\_without\_children \to men\_without\_children}$("motherhood") = (0; -0.54; 0.23) | 0.59 | $WT_{women\_with\_children \to men\_without\_children}$("fatherhood") = (-0.38; -0.52; -0.77) | 1 |
| $WT_{women\_without\_children \to women\_with\_children}$("pregnancy") = (0.74; -0.29; 0.56) | 0.97 | $WT_{women\_without\_children \to men\_without\_children}$("infant") = (-0.23; -0.54; 0) | 0.59 | $WT_{women\_with\_children \to men\_without\_children}$("sole custody of child") = (-0.62; 0.3; -0.69) | 0.97 |
| $WT_{women\_without\_children \to women\_with\_children}$("preemie") = (0.67; -0.16; 0.43) | 0.81 | $WT_{women\_without\_children \to men\_without\_children}$("intimate relationship") = (-0.46; 0.23; -0.08) | 0.52 | $WT_{women\_with\_children \to men\_without\_children}$("artificial fertilization") = (-0.48; 0.69; -0.49) | 0.97 |
| $WT_{women\_without\_children \to women\_with\_children}$("baby colic") = (0.33; -0.67; 0.01) | 0.75 | $WT_{women\_without\_children \to men\_without\_children}$("artificial fertilization") = (0.23; -0.23; -0.31) | 0.45 | $WT_{women\_with\_children \to men\_without\_children}$("intimate relationship") = (-0.28; -0.1; -0.67) | 0.73 |
| $WT_{women\_without\_children \to women\_with\_children}$("intimate relationship") = (-0.18; 0.33; 0.59) | 0.7 | $WT_{women\_without\_children \to men\_without\_children}$("fatherhood") = (0.31; -0.23; -0.08) | 0.39 | $WT_{women\_with\_children \to men\_without\_children}$("abortion") = (0.48; 0.18; 0.46) | 0.69 |
| $WT_{women\_without\_children \to women\_with\_children}$("breastfeeding") = (0.44; -0.29; 0.17) | 0.55 | $WT_{women\_without\_children \to men\_without\_children}$("abortion") = (0.38; -0.08; 0) | 0.39 | $WT_{women\_with\_children \to men\_without\_children}$("preemie") = (-0.44; -0.46; -0.2) | 0.67 |
| $WT_{women\_without\_children \to women\_with\_children}$("abortion") = (-0.1; -0.26; -0.46) | 0.54 | $WT_{women\_without\_children \to men\_without\_children}$("miscarriage") = (0.3; 0.07; 0.23) | 0.38 | $WT_{women\_with\_children \to men\_without\_children}$("breastfeeding") = (-0.59; 0.13; -0.17) | 0.63 |
| $WT_{women\_without\_children \to women\_with\_children}$("childlessness") = (-0.19; 0.24; -0.24) | 0.39 | $WT_{women\_without\_children \to men\_without\_children}$("breastfeeding") = (-0.15; -0.16; 0) | 0.22 | $WT_{women\_with\_children \to men\_without\_children}$("baby colic") = (0.05; 0.29; 0.53) | 0.61 |
| $WT_{women\_without\_children \to women\_with\_children}$("sexuality") = (-0.33; -0.14; 0.09) | 0.37 | $WT_{women\_without\_children \to men\_without\_children}$("sexuality") = (0.16; 0; -0.15) | 0.22 | $WT_{women\_with\_children \to men\_without\_children}$("sexuality") = (0.49; 0.14; -0.24) | 0.56 |
| $WT_{women\_without\_children \rightarrow women\_with\_children}$("miscarriage") = (-0.08; 0.28; -0.08) | 0.3 | $WT_{women\_without\_children \rightarrow men\_without\_children}$("sole custody of child") = (0.15; 0; -0.15) | 0.21 | $WT_{women\_with\_children \rightarrow men\_without\_children}$("miscarriage") = (0.38; -0.21; 0.31) | 0.53 |

## Some cluster-based transformation vectors for gender-parental subgroups:

$CT_{women\_without\_children \rightarrow women\_with\_children}$(cluster A) = (0.074; -0.169; 0.129)
$CT_{women\_without\_children \rightarrow women\_with\_children}$(cluster B) = (-0.013; -0.017; -0.099)
$CT_{women\_without\_children \rightarrow women\_with\_children}$(cluster C) = (-0.087; -0.035; -0.103)
$CT_{women\_without\_children \rightarrow women\_with\_children}$(cluster D) = (0.091; -0.108; 0.087)
$CT_{women\_without\_children \rightarrow women\_with\_children}$(cluster E) = (-0.099; -0.028; -0.044)

$CT_{women\_without\_children \rightarrow men\_without\_children}$(cluster A) = (-0.016; -0.085; -0.111)
$CT_{women\_without\_children \rightarrow men\_without\_children}$(cluster B) = (0.055; -0.169; -0.029)
$CT_{women\_without\_children \rightarrow men\_without\_children}$(cluster C) = (-0.025; -0.045; 0.051)
$CT_{women\_without\_children \rightarrow men\_without\_children}$(cluster D) = (-0.094; -0.309; -0.068)
$CT_{women\_without\_children \rightarrow men\_without\_children}$(cluster E) = (0.062; -0.19; 0.0680)

$CT_{women\_with\_children \rightarrow men\_without\_children}$(cluster A) = (-0.090; 0.084; -0.24)
$CT_{women\_with\_children \rightarrow men\_without\_children}$(cluster B) = (0.068; -0.152; 0.070)
$CT_{women\_with\_children \rightarrow men\_without\_children}$(cluster C) = (0.062; -0.01; 0.154)
$CT_{women\_with\_children \rightarrow men\_without\_children}$(cluster D) = (-0.185; -0.201; -0.155)
$CT_{women\_with\_children \rightarrow men\_without\_children}$(cluster E) = (0.161; -0.162; 0.112)

# Appendix E

**Table E1.** The number of average vectors of emotional adjectives for each subgroup that have a direction pointing into one of the eight corners of the emotional vector space. a) The values of pleasure, arousal and dominance are more or less than zero and b) these values are more than 1 or less than -1. Notation: w = women, m = men, wwc = women with children, wwoc = women without children, mwc = men with children and mwoc = men without children.

a) Eight corners of the emotional vector space with values more or less than zero

|  | group | dominance <0 | | dominance >0 | |
|---|---|---|---|---|---|
|  |  | arousal <0 | arousal >0 | arousal <0 | arousal >0 |
| pleasure <0 | all | 41 | 57 | 0 | 0 |
|  | w | 39 | 56 | 0 | 0 |
|  | m | 39 | 55 | 0 | 0 |
|  | wwc | 49 | 42 | 0 | 0 |
|  | wwoc | 38 | 57 | 0 | 0 |
|  | mwc | 22 | 57 | 0 | 0 |
|  | mwoc | 41 | 52 | 0 | 0 |
| pleasure >0 | all | 1 | 4 | 69 | 23 |
|  | w | 4 | 1 | 67 | 23 |
|  | m | 0 | 4 | 62 | 25 |
|  | wwc | 2 | 1 | 73 | 12 |
|  | wwoc | 1 | 3 | 63 | 26 |
|  | mwc | 2 | 4 | 19 | 36 |
|  | mwoc | 0 | 3 | 63 | 23 |

b) Eight corners of the emotional vector space with values more than 1 or less than -1

|  | group | dominance <-1 | | dominance >1 | |
|---|---|---|---|---|---|
|  |  | arousal <-1 | arousal >1 | arousal <-1 | arousal >1 |
| pleasure <-1 | all | 4 | 3 | 0 | 0 |
|  | w | 5 | 8 | 0 | 0 |
|  | m | 2 | 1 | 0 | 0 |
|  | wwc | 4 | 7 | 0 | 0 |
|  | wwoc | 6 | 12 | 0 | 0 |
|  | mwc | 0 | 2 | 0 | 0 |
|  | mwoc | 2 | 1 | 0 | 0 |
| pleasure >1 | all | 0 | 0 | 1 | 0 |
|  | w | 0 | 0 | 3 | 0 |
|  | m | 0 | 0 | 0 | 0 |
|  | wwc | 0 | 0 | 4 | 0 |
|  | wwoc | 0 | 0 | 4 | 1 |
|  | mwc | 0 | 0 | 0 | 0 |
|  | mwoc | 0 | 0 | 0 | 0 |

**Table E2.** Some distinctive average vectors of emotional adjectives towards the eight corners of the emotional vector space with affective rating values more than 1 or less than -1, in six pair-wise comparison combinations between three gender-parental subgroups Women with children (wwc; n=8), Women without children (wwoc; n=13) and Men without children (mwoc; n=13). Emotional adjectives are listed in the decreasing order of distance (shown in parenthesis).

| Eight corners of the emotional vector space with affective rating values more than 1 or less than -1 | | | | |
|---|---|---|---|---|
| | | dominance <-1 | | dominance >1 |
| | group | arousal <-1 | arousal >1 | arousal <-1 | arousal >1 |
| pleasure <-1 | wwc & ¬wwoc: | melancholic (0.71) weak (0.6) | mistreated (0.83) disoriented (0.64) | No value | No value |
| | ¬wwc & wwoc: | empty (1.02) powerless (0.93) sad (0.88) | nervous (0.99) abandoned (0.69) desperate (0.66) | No value | No value |
| | wwoc & ¬mwoc: | powerless (1.22) sad (1.12) depressed (0.84) | desperate (1.44) abandoned (1.29) shocked (1.26) | No value | No value |
| | ¬wwoc & mwoc: | No value | No value | No value | No value |
| | wwc & ¬mwoc: | depressed (1.19) weak (0.95) melancholic (0.57) | shocked (1.28) disoriented (1.13) fearful (1.02) | No value | No value |
| | ¬wwc & mwoc: | empty (0.97) | No value | No value | No value |
| pleasure >1 | wwc & ¬wwoc: | No value | No value | relaxed (0.34) calm (0.34) | No value |
| | ¬wwc & wwoc: | No value | No value | whole (0.74) cohesive (0.55) | passionate (0.33) |
| | wwoc & ¬mwoc: | No value | No value | cohesive (1.1) whole (0.98) balanced (0.81) | passionate (0.83) |
| | ¬wwoc & mwoc: | No value | No value | No value | No value |
| | wwc & ¬mwoc: | No value | No value | balanced (0.92) relaxed (0.84) calm (0.59) | No value |
| | ¬wwc & mwoc: | No value | No value | No value | No value |

## Some lists of extreme affective shift models for gender-parental subgroups:

L (pleasure <0 & arousal <0 & dominance <0; wwc & ¬wwoc) = { uncomfortable (1.18); shared sense of shame (1.096); awkward (1.056); skeptical (0.999); helpless (0.959); distrustful (0.874); remorseful (0.836); concerned (0.747); embarrassed (0.714); bothered (0.713); stupid (0.627); irresolute (0.616); gloating (0.601); frustrated (0.579); reserved (0.461); disapproving (0.268); }
L (pleasure <0 & arousal <0 & dominance <0; ¬wwc & wwoc) = { small (1.012); disappointed (0.874); negative (0.743); inadequate (0.646); stuck (0.567); dissatisfied (0.379); }
L (pleasure <0 & arousal <0 & dominance <0; wwoc & ¬mwoc) = { burned out (1.542); disappointed (1.271); stuck (0.913); inadequate (0.803); negative (0.777); useless (0.679); }
L (pleasure <0 & arousal <0 & dominance <0; ¬wwoc & mwoc) = { lost (1.791); insecure (1.058); irresolute (1.049); reserved (0.792); worthless (0.773); remorseful (0.773); embarrassed (0.653); stupid (0.644); disapproving (0.361); }
L (pleasure <0 & arousal <0 & dominance <0; wwc & ¬mwoc) = { gloating (1.15); burned out (1.008); helpless (0.954); awkward (0.921); frustrated (0.891); distrustful (0.837); skeptical (0.821); concerned (0.748); shared sense of shame (0.748); bothered (0.748); useless (0.738); uncomfortable (0.488); }
L (pleasure <0 & arousal <0 & dominance <0; ¬wwc & mwoc) = { small (1.19); dissatisfied (0.831); lost (0.811); insecure (0.486); }

L (pleasure <0 & arousal >0 & dominance <0; wwc & ¬wwoc) = { disappointed (0.874); offended (0.777); inadequate (0.646); stuck (0.567); dissatisfied (0.379); }
L (pleasure <0 & arousal >0 & dominance <0; ¬wwc & wwoc) = { aggressive (1.185); uncomfortable (1.18); shared sense of shame (1.096); awkward (1.056); skeptical (0.999); helpless (0.959); anxious (0.916); distrustful (0.874); worthless (0.761); concerned (0.747); embarrassed (0.714); bothered (0.713); stupid (0.627); irresolute (0.616); gloating (0.601); frustrated (0.579); judgmental (0.493); reserved (0.461); envious (0.359); disapproving (0.268); }
L (pleasure <0 & arousal >0 & dominance <0; wwoc & ¬mwoc) = { lost (1.791); aggressive (1.573); skeptical (1.214); anxious (1.146); insecure (1.058); irresolute (1.049); terrible (0.828); reserved (0.792); worthless (0.773); embarrassed (0.653); stupid (0.644); disapproving (0.361); }
L (pleasure <0 & arousal >0 & dominance <0; ¬wwoc & mwoc) = { burned out (1.542); disappointed (1.271); stuck (0.913); offended (0.874); inadequate (0.803); negative (0.777); overwhelmed (0.474); }
L (pleasure <0 & arousal >0 & dominance <0; wwc & ¬mwoc) = { terrible (1.469); dissatisfied (0.831); lost (0.811); insecure (0.486); }
L (pleasure <0 & arousal >0 & dominance <0; ¬wwc & mwoc) = { gloating (1.15); burned out (1.008); judgmental (0.999); helpless (0.954); awkward (0.921); frustrated (0.891); distrustful (0.837); concerned (0.748); shared sense of shame (0.748); bothered (0.748); envious (0.674); negative (0.603); uncomfortable (0.488); overwhelmed (0.487); }

L (pleasure <0 & arousal <0 & dominance >0; wwc & ¬wwoc) = { No value; }
L (pleasure <0 & arousal <0 & dominance >0; ¬wwc & wwoc) = { No value; }
L (pleasure <0 & arousal <0 & dominance >0; wwoc & ¬mwoc) = { No value; }
L (pleasure <0 & arousal <0 & dominance >0; ¬wwoc & mwoc) = { No value; }

L (pleasure <0 & arousal <0 & dominance >0; wwc & ¬mwoc) = { No value; }
L (pleasure <0 & arousal <0 & dominance >0; ¬wwc & mwoc) = { No value; }

L (pleasure <0 & arousal >0 & dominance >0; wwc & ¬wwoc) = { No value; }
L (pleasure <0 & arousal >0 & dominance >0; ¬wwc & wwoc) = { No value; }
L (pleasure <0 & arousal >0 & dominance >0; wwoc & ¬mwoc) = { No value; }
L (pleasure <0 & arousal >0 & dominance >0; ¬wwoc & mwoc) = { No value; }
L (pleasure <0 & arousal >0 & dominance >0; wwc & ¬mwoc) = { No value; }
L (pleasure <0 & arousal >0 & dominance >0; ¬wwc & mwoc) = { No value; }

L (pleasure >0 & arousal <0 & dominance <0; wwc & ¬wwoc) = { astonished (0.878); expecting (0.604); }
L (pleasure >0 & arousal <0 & dominance <0; ¬wwc & wwoc) = { kind (0.515); }
L (pleasure >0 & arousal <0 & dominance <0; wwoc & ¬mwoc) = { kind (0.853); }
L (pleasure >0 & arousal <0 & dominance <0; ¬wwoc & mwoc) = { No value; }
L (pleasure >0 & arousal <0 & dominance <0; wwc & ¬mwoc) = { astonished (1.084); expecting (0.5); }
L (pleasure >0 & arousal <0 & dominance <0; ¬wwc & mwoc) = { No value; }

L (pleasure >0 & arousal >0 & dominance <0; wwc & ¬wwoc) = { touched (0.406); }
L (pleasure >0 & arousal >0 & dominance <0; ¬wwc & wwoc) = { naked (1.533); astonished (0.878); blissful (0.636); }
L (pleasure >0 & arousal >0 & dominance <0; wwoc & ¬mwoc) = { naked (1.451); astonished (0.493); }
L (pleasure >0 & arousal >0 & dominance <0; ¬wwoc & mwoc) = { ecstatic (0.843); touched (0.455); }
L (pleasure >0 & arousal >0 & dominance <0; wwc & ¬mwoc) = { No value; }
L (pleasure >0 & arousal >0 & dominance <0; ¬wwc & mwoc) = { ecstatic (1.254); blissful (0.864); }

L (pleasure >0 & arousal <0 & dominance >0; wwc & ¬wwoc) = { ecstatic (1.017); small (1.012); alert (0.944); admiring (0.877); fantastic (0.675); hopeful (0.645); powerful (0.61); kind (0.515); capable (0.493); courageous (0.46); gritty (0.368); elevated (0.276); }
L (pleasure >0 & arousal <0 & dominance >0; ¬wwc & wwoc) = { productive (0.733); happy (0.592); }
L (pleasure >0 & arousal <0 & dominance >0; wwoc & ¬mwoc) = { special (1.527); interested (0.955); lovely (0.821); lively (0.761); delighted (0.761); amused (0.639); shared sense of pride (0.625); funny (0.56); }
L (pleasure >0 & arousal <0 & dominance >0; ¬wwoc & mwoc) = { expecting (0.88); capable (0.86); kind (0.853); hopeful (0.644); prolific (0.576); amazing (0.522); powerful (0.474); elevated (0.231); }
L (pleasure >0 & arousal <0 & dominance >0; wwc & ¬mwoc) = { ecstatic (1.254); special (1.224); small (1.19); fantastic (1.164); amused (1.077); interested (1.058); admiring (0.987); lively (0.892); alert (0.859); delighted (0.797); courageous (0.718); lovely (0.702); shared sense of pride (0.697); funny (0.58); gritty (0.477); }
L (pleasure >0 & arousal <0 & dominance >0; ¬wwc & mwoc) = { productive (1.085); happy (0.598); expecting (0.5); amazing (0.381); prolific (0.237); }

L (pleasure >0 & arousal >0 & dominance >0; wwc & ¬wwoc) = { productive (0.733); blissful (0.636); }
L (pleasure >0 & arousal >0 & dominance >0; ¬wwc & wwoc) = { euphoric (1.735); ecstatic (1.017); alert (0.944); desired (0.771); energetic (0.758); fantastic (0.675); cheerful (0.661); hopeful (0.645); motivated (0.613); powerful (0.61); capable (0.493); curious (0.484); courageous (0.46); surprised (0.404); gritty (0.368); prolific (0.352); }
L (pleasure >0 & arousal >0 & dominance >0; wwoc & ¬mwoc) = { capable (0.86); cheerful (0.85); ecstatic (0.843); efficient (0.773); wonderful (0.675); hopeful (0.644); prolific (0.576); amazing (0.522); powerful (0.474); }
L (pleasure >0 & arousal >0 & dominance >0; ¬wwoc & mwoc) = { naked (1.451); interested (0.955); lovely (0.821); lively (0.761); shared sense of pride (0.625); admiring (0.606); funny (0.56); astonished (0.493); }
L (pleasure >0 & arousal >0 & dominance >0; wwc & ¬mwoc) = { productive (1.085); blissful (0.864); efficient (0.635); competitive (0.543); amazing (0.381); wonderful (0.325); }
L (pleasure >0 & arousal >0 & dominance >0; ¬wwc & mwoc) = { naked (2.037); euphoric (1.656); energetic (1.238); fantastic (1.164); astonished (1.084); interested (1.058); admiring (0.987); lively (0.892); alert (0.859); desired (0.784); courageous (0.718); lovely (0.702); shared sense of pride (0.697); surprised (0.586); motivated (0.583); funny (0.58); gritty (0.477); curious (0.378); }

# Appendix F

**Table F1.** For each of five clusters of a gender-parental subgroup those emotional adjectives that have exactly or nearly the same direction as the average vector of a cluster (based on a threshold of cosine similarity value of at least 0.99). Notation: emot. adj. = emotional adjective, distance from the avg vector of the cluster = distance of the average vector of the emotional adjective from the average vector of the cluster, distance from the origo = distance of the average vector of the emotional adjective from the origo, cosine btw the avg vectors of the cluster and the emot. adj. = cosine similarity measure between the average vector of the cluster and the average vector of the emotional adjective, a prefix > indicates three emotional adjectives having the shortest distance to the average vector of a cluster (listed also in Table 5 for the cluster in question), a suffix @ indicates emotional adjectives closest to the average vector of a cluster even if they are not having a cosine similarity value of at least 0.99.

| Women without children (n=13) | | | | Women with children (n=8) | | | | Men without children (n=13) | | | |
|---|---|---|---|---|---|---|---|---|---|---|---|
| emot. adj. | distance from the avg vector of the cluster | distance from the origo | cosine btw the avg vectors of the cluster and the emot. adj. | emot. adj. | distance from the avg vector of the cluster | distance from the origo | cosine btw the avg vectors of the cluster and the emot. adj. | emot. adj. | distance from the avg vector of the cluster | distance from the origo | cosine btw the avg vectors of the cluster and the emot. adj. |
| **A** | | | | **A** | | | | **A** | | | |
| > admiring @ | 0.24 | 1.41 | 0.99 | > powerful | 0.2 | 1.64 | 0.99 | shared sense of pride | 0.4 | | 1 |
| > curious | 0.15 | 1.46 | 0.99 | > proud | 0.24 | 1.71 | 0.99 | curious | 0.4 | 1.01 | 1 |
| > gritty | 0.31 | 1.72 | 0.99 | > wonderful | 0.23 | 1.81 | 1 | wow! | 0.38 | 1.03 | 0.99 |
| hopeful | 0.33 | 1.72 | 0.99 | energetic | 0.34 | 1.91 | 1 | > funny | 0.01 | 1.39 | 1 |
| amazing | 0.39 | 1.8 | 0.99 | gritty | 0.38 | 1.91 | 0.99 | > proud @ | 0.24 | 1.41 | 0.99 |
| superb | 0.5 | 1.93 | 0.99 | desired | 0.43 | 1.96 | 0.99 | > motivated | 0.33 | 1.71 | 1 |
| fantastic | 0.63 | 2.07 | 0.99 | amazing | 0.4 | 1.98 | 1 | wonderful | 0.51 | 1.86 | 0.99 |
| | | | | loving | 0.69 | 2.23 | 0.99 | | | | |
| | | | | productive | 0.77 | 2.34 | 1 | | | | |
| | | | | happy | 0.85 | 2.43 | 1 | | | | |
| **B** | | | | **B** | | | | **B** | | | |
| > irritated | 0.27 | 1.59 | 0.99 | pessimistic | 0.96 | 0.88 | 1 | confused | 1.29 | 0.53 | 0.99 |
| > miserable_I @ | 0.29 | 1.86 | 0.99 | remorseful | 0.57 | 1.31 | 0.99 | sorry | 0.96 | 0.87 | 1 |
| > disappointed @ | 0.12 | 1.88 | 1 | > bothered @ | 0.35 | 1.64 | 0.99 | indifferent | 0.7 | 1.13 | 1 |
| lonely | 0.56 | 2.32 | 1 | > lonely @ | 0.34 | 1.81 | 1 | small | 0.68 | 1.16 | 0.99 |
| sad | 0.71 | 2.42 | 0.99 | > tired @ | 0.35 | 1.91 | 0.98 | pessimistic | 0.61 | 1.22 | 1 |
| burned out | 0.79 | 2.54 | 1 | miserable_II | 0.37 | 2.13 | 0.99 | > melancholic @ | 0.26 | 1.82 | 0.99 |
| miserable_II | 0.84 | 2.56 | 0.99 | hopeless | 0.51 | 2.34 | 1 | > exhausted | 0.14 | 1.84 | 1 |
| depressed | 0.86 | 2.61 | 1 | numb | 0.7 | 2.51 | 1 | > drained | 0.09 | 1.89 | 1 |
| hopeless | 1.02 | 2.75 | 0.99 | insignificant | 1.05 | 2.84 | 0.99 | | | | |
| insignificant | 1.15 | 2.9 | 1 | depressed | 1.26 | 3.09 | 1 | | | | |
| powerless | 1.23 | 2.96 | 0.99 | | | | | | | | |
| **C** | | | | **C** | | | | **C** | | | |
| shared sense of pride | 0.33 | 1.3 | 1 | optimistic | 0.28 | 1.31 | 0.99 | appreciative | 0.77 | 0.86 | 1 |
| compassionate | 0.28 | 1.39 | 0.99 | > understanding | 0.14 | 1.49 | 1 | blessed | 0.74 | 0.9 | 1 |
| > positive | 0.18 | 1.54 | 0.99 | > compassionate | 0.06 | 1.55 | 1 | respectful | 0.34 | 1.35 | 0.99 |
| > considerate | 0.17 | 1.6 | 0.99 | special | 0.22 | 1.56 | 0.99 | open-minded | 0.27 | 1.45 | 0.99 |
| > committed | 0.1 | 1.63 | 1 | respected | 0.22 | 1.56 | 0.99 | caring | 0.19 | 1.46 | 1 |
| affectionate | 0.22 | 1.72 | 0.99 | > positive | 0.19 | 1.59 | 0.99 | > whole | 0.12 | 1.53 | 1 |
| present | 0.3 | 1.8 | 1 | present | 0.2 | 1.65 | 1 | > harmonious | 0.1 | 1.55 | 1 |
| loving | 0.37 | 1.9 | 0.99 | assured | 0.27 | 1.76 | 1 | committed | 0.17 | 1.57 | 1 |
| respectful | 0.39 | 1.98 | 1 | warm | 0.3 | 1.77 | 1 | understanding | 0.23 | 1.63 | 0.99 |
| appreciative | 0.44 | 2.01 | 1 | open-minded | 0.3 | 1.79 | 1 | > present | 0.13 | 1.7 | 1 |
| satisfied | 0.42 | 2.02 | 1 | accepting | 0.51 | 1.98 | 0.99 | competent | 0.24 | 1.81 | 1 |
| cohesive | 0.63 | 2.19 | 0.99 | unique | 0.71 | 2.22 | 1 | independent | 0.42 | 1.99 | 0.99 |
| | | | | appreciative | 0.75 | 2.25 | 1 | | | | |
| **D** | | | | **D** | | | | **D** | | | |
| > gloating | 0.17 | 1.46 | 0.99 | > tense_II | 0.15 | 1.26 | 1 | longing | 0.9 | 0.71 | 1 |
| > disapproving | 0.14 | 1.48 | 1 | > lost | 0.24 | 1.51 | 0.99 | disapproving | 0.47 | 1.16 | 1 |
| > bothered | 0.12 | 1.51 | 1 | > disappointed | 0.24 | 1.51 | 1 | terrible | 0.38 | 1.25 | 1 |
| embarrassed | 0.32 | 1.78 | 0.99 | insecure | 0.29 | 1.61 | 0.99 | disgusting | 0.35 | 1.31 | 0.99 |
| frustrated | 0.34 | 1.8 | 0.99 | negative | 0.37 | 1.7 | 1 | > stupid | 0.22 | 1.44 | 1 |
| stupid | 0.43 | 1.9 | 1 | dissatisfied | 0.42 | 1.77 | 1 | > miserable_II | 0.18 | 1.65 | 1 |
| guilty | 0.48 | 1.98 | 1 | judgmental | 0.6 | 1.95 | 0.99 | > ashamed | 0.2 | 1.7 | 0.99 |
| envious | 0.59 | 2.09 | 1 | anxious | 0.59 | 1.95 | 1 | hopeless | 0.23 | 1.82 | 1 |
| mistreated | 0.91 | 2.42 | 1 | inadequate | 0.65 | 1.98 | 1 | worthless | 0.29 | 1.88 | 1 |
| helpless | 0.93 | 2.42 | 1 | nervous | 0.65 | 1.99 | 1 | depressed | 0.47 | 2.03 | 0.99 |
| worthless | 0.96 | 2.46 | 1 | envious | 0.77 | 2.13 | 1 | useless | 0.54 | 2.14 | 1 |
| hurt | 0.97 | 2.47 | 1 | disgusting | 0.8 | 2.15 | 1 | | | | |
| oppressed | 1.13 | 2.6 | 0.99 | oppressed | 1.37 | 2.74 | 1 | | | | |
| forced | 1.43 | 2.91 | 0.99 | | | | | | | | |
| **E** | | | | **E** | | | | **E** | | | |
| skeptical | 0.44 | 1.62 | 0.99 | annoyed | 0.68 | 1.42 | 1 | shared sense of shame | 1.01 | 0.85 | 1 |
| concerned | 0.38 | 1.72 | 0.99 | guilty | 0.31 | 2.22 | 0.99 | tense_II | 0.81 | 1.05 | 1 |
| disoriented | 0.37 | 1.72 | 0.99 | disoriented | 0.34 | 2.33 | 0.99 | manic | 0.75 | 1.12 | 0.99 |
| > uncomfortable | 0.18 | 1.87 | 1 | offended | 0.35 | 2.34 | 0.99 | annoyed | 0.53 | 1.33 | 1 |
| > tense_II | 0.3 | 1.88 | 0.99 | > hurt | 0.28 | 2.35 | 1 | stressed | 0.43 | 1.48 | 0.99 |
| jealous | 0.37 | 2.27 | 0.99 | fearful | 0.49 | 2.49 | 0.99 | jealous | 0.25 | 1.6 | 1 |
| > ashamed | 0.34 | 2.33 | 1 | mistreated | 0.49 | 2.57 | 1 | > shocked @ | 0.31 | 1.72 | 0.99 |
| stressed | 0.44 | 2.36 | 0.99 | upset | 0.51 | 2.6 | 1 | mistreated | 0.13 | 1.94 | 1 |
| anxious | 0.48 | 2.42 | 0.99 | alarmed | 0.57 | 2.62 | 1 | offended | 0.46 | 2.28 | 1 |
| upset | 0.54 | 2.52 | 1 | abused | 0.99 | 3.04 | 0.99 | | | | |
| betrayed | 0.68 | 2.68 | 1 | panicked | 1 | 3.05 | 0.99 | | | | |
| desperate | 0.85 | 2.84 | 1 | | | | | | | | |
| alarmed | 0.94 | 2.9 | 0.99 | | | | | | | | |

Some lists of scale-cluster models for gender-parental subgroups:

Notation: a prefix > indicates three emotional adjectives having the shortest distance to the average vector of a cluster (listed also in Table 5 for the cluster in question), a suffix @ indicates emotional adjectives closest to the average vector of a cluster even if they are not having a cosine similarity value of at least 0.99.

SL (women without children; cluster A) = { "> admiring @" (1.41); "> curious" (1.46); "> gritty" (1.72); "hopeful" (1.72); "amazing" (1.8); "superb" (1.93); "fantastic" (2.07); }
SL (women without children; cluster B) = { "> irritated" (1.59); "> miserable_I @" (1.86); "> disappointed" (1.88); "lonely" (2.32); "sad" (2.42); "burned out" (2.54); "miserable_II" (2.56); "depressed" (2.61); "hopeless" (2.75); "insignificant" (2.9); "powerless" (2.96); }
SL (women without children; cluster C) = { "shared sense of pride" (1.3); "compassionate" (1.39); "> positive" (1.54); "> considerate" (1.6); "> committed" (1.63); "affectionate" (1.72); "present" (1.8); "loving" (1.9); "respectful" (1.98); "appreciative" (2.01); "satisfied" (2.02); "cohesive" (2.19); }
SL (women without children; cluster D) = { "> gloating" (1.46); "> disapproving" (1.48); "> bothered" (1.51); "embarrassed" (1.78); "frustrated" (1.8); "stupid" (1.9); "guilty" (1.98); "envious" (2.09); "mistreated" (2.42); "helpless" (2.42); "worthless" (2.46); "hurt" (2.47); "oppressed" (2.6); "forced" (2.91); }
SL (women without children; cluster E) = { "skeptical" (1.62); "concerned" (1.72); "disoriented" (1.72); "> uncomfortable" (1.87); "> tense_II" (1.88); "jealous" (2.27); "> ashamed" (2.33); "stressed" (2.36); "anxious" (2.42); "upset" (2.52); "betrayed" (2.68); "desperate" (2.84); "alarmed" (2.9); }

SL (women with children; cluster A) = { "> powerful" (1.64); "> proud" (1.71); "> wonderful" (1.81); "energetic" (1.91); "gritty" (1.91); "desired" (1.96); "amazing" (1.98); "loving" (2.23); "productive" (2.34); "happy" (2.43); }
SL (women with children; cluster B) = { "pessimistic" (0.88); "remorseful" (1.31); "> bothered @" (1.64); "> lonely @" (1.81); "> tired @" (1.91); "miserable_II" (2.13); "hopeless" (2.34); "numb" (2.51); "insignificant" (2.84); "depressed" (3.09); }
SL (women with children; cluster C) = { "optimistic" (1.31); "> understanding" (1.49); "> compassionate" (1.55); "special" (1.56); "respected" (1.56); "> positive" (1.59); "present" (1.65); "close" (1.7); "assured" (1.76); "warm" (1.77); "open-minded" (1.79); "accepting" (1.96); "unique" (2.22); "appreciative" (2.25); }
SL (women with children; cluster D) = { "> tense_II" (1.26); "> lost" (1.51); "> disappointed" (1.51); "insecure" (1.61); "negative" (1.7); "dissatisfied" (1.77); "judgmental" (1.95); "anxious" (1.95); "inadequate" (1.98); "nervous" (1.99); "envious" (2.13); "disgusting" (2.15); "oppressed" (2.74); }

SL (women with children; cluster E) = { "annoyed" (1.42); "> guilty" (2.22); "> disoriented" (2.33); "offended" (2.34); "> hurt" (2.35); "fearful" (2.49); "mistreated" (2.57); "upset" (2.6); "alarmed" (2.62); "abused" (3.04); "panicked" (3.05); }
SL (men without children; cluster A) = { "shared sense of pride" (1); "curious" (1.01); "wow!" (1.03); "> funny" (1.39); "> proud @" (1.41); "> motivated" (1.71); "wonderful" (1.86); }
SL (men without children; cluster B) = { "confused" (0.53); "sorry" (0.87); "indifferent" (1.13); "small" (1.16); "pessimistic" (1.22); "> melancholic @" (1.82); "> exhausted" (1.84); "> drained" (1.89); }
SL (men without children; cluster C) = { "appreciative" (0.86); "blessed" (0.9); "respectful" (1.35); "open-minded" (1.45); "caring" (1.46); "> whole" (1.53); "> harmonious" (1.55); "committed" (1.57); "understanding" (1.63); "> present" (1.7); "competent" (1.81); "independent" (1.99); }
SL (men without children; cluster D) = { "longing" (0.71); "disapproving" (1.16); "terrible" (1.25); "disgusting" (1.31); "> stupid" (1.44); "> miserable_II" (1.65); "> ashamed" (1.72); "hopeless" (1.82); "worthless" (1.88); "hurt" (2.03); "depressed" (2.05); "useless" (2.14); }
SL (men without children; cluster E) = { "shared sense of shame" (0.85); "tense_II" (1.05); "manic" (1.12); "annoyed" (1.33); "stressed" (1.48); "jealous" (1.6); "> shocked @" (1.72); "> mistreated" (1.94); "offended" (2.28); }

Some lists of the attraction-cluster model for gender-parental subgroups showing the nearest average vectors of clusters in an ascending order in respect to the distance between the average vector of the pregnancy-related noun and the average vector of the cluster for the subgroup in question:

AL (women without children; "intimate relationship") = { C, A, D, B, E }
AL (women without children; "motherhood") = { A, C, D, B, E }
AL (women without children; "fatherhood") = { C, A, D, B, E }
AL (women without children; "infant") = { A, C, D, B, E }
AL (women without children; "fetus") = { D, E, B, A, C }
AL (women without children; "pregnancy") = { A, D, C, E, B }
AL (women without children; "giving birth") = { D, E, B, A, C }
AL (women without children; "breastfeeding") = { C, A, D, B, E }
AL (women without children; "baby colic") = { E, D, B, A, C }
AL (women without children; "miscarriage") = { E, D, B, A, C }
AL (women without children; "abortion") = { D, E, B, A, C }
AL (women without children; "preemie") = { D, E, B, A, C }

AL (women without children; "childlessness") = { B, D, E, C, A }
AL (women without children; "sexuality") = { A, C, D, B, E }
AL (women without children; "sole custody of child") = { D, B, E, A, C }
AL (women without children; "artificial fertilization") = { D, B, A, C, E }

AL (women with children; "intimate relationship") = { A, C, D, B, E }
AL (women with children; "motherhood") = { A, C, D, B, E }
AL (women with children; "fatherhood") = { A, C, D, B, E }
AL (women with children; "infant") = { A, C, D, B, E }
AL (women with children; "fetus") = { A, C, D, B, E }
AL (women with children; "pregnancy") = { A, C, D, B, E }
AL (women with children; "giving birth") = { A, D, E, C, B }
AL (women with children; "breastfeeding") = { C, A, D, B, E }
AL (women with children; "baby colic") = { D, E, B, C, A }
AL (women with children; "miscarriage") = { E, D, B, A, C }
AL (women with children; "abortion") = { D, E, B, C, A }
AL (women with children; "preemie") = { D, E, B, C, A }
AL (women with children; "childlessness") = { B, D, E, C, A }
AL (women with children; "sexuality") = { A, C, D, B, E }
AL (women with children; "sole custody of child") = { C, D, A, B, E }
AL (women with children; "artificial fertilization") = { C, A, D, B, E }

AL (men without children; "intimate relationship") = { A, C, D, B, E }
AL (men without children; "motherhood") = { A, C, D, B, E }
AL (men without children; "fatherhood") = { C, A, B, D, E }
AL (men without children; "infant") = { C, A, B, D, E }
AL (men without children; "fetus") = { B, D, A, C, E }
AL (men without children; "pregnancy") = { A, C, D, B, E }
AL (men without children; "giving birth") = { D, E, B, A, C }
AL (men without children; "breastfeeding") = { C, A, B, D, E }
AL (men without children; "baby colic") = { E, D, B, A, C }
AL (men without children; "miscarriage") = { E, D, B, A, C }
AL (men without children; "abortion") = { D, E, B, A, C }
AL (men without children; "preemie") = { D, B, E, A, C }
AL (men without children; "childlessness") = { D, B, E, C, A }
AL (men without children; "sexuality") = { A, C, D, E, B }
AL (men without children; "sole custody of child") = { D, B, E, A, C }
AL (men without children; "artificial fertilization") = { D, B, A, C, E }

zzzzzzz